%% file: 2916.tex
\documentclass[twocolumn]{aa}
\usepackage{graphicx}
\usepackage{natbib}
\newcommand{\teff}{$T_{\rm eff}$}

\newcommand{\nodata}{\ldots}

\begin{document}
   \title{Abundances in giant stars of the globular cluster 
   NGC 6752\thanks{Based on observations obtained with 
   the ESO Very Large Telescope UVES spectrograph for programmes 67.D-0145 
   and 65.L-0165(A)}}

\authorrunning{Yong et al.}
\titlerunning{Abundances in giant stars in NGC 6752}

   \author{D. Yong
          \inst{1,2}
          \and
          F. Grundahl\inst{3}
          \and
          P. E. Nissen\inst{3}
          \and 
          H. R. Jensen\inst{3}
          \and
          D. L. Lambert\inst{1}
          }


   \institute{Department of Astronomy, University of Texas, Austin, TX 78712, USA \\
   \email{tofu,dll@astro.as.utexas.edu}
   \and
   Department of Physics and Astronomy, University of North Carolina, NC 27599, USA
   \and
   Institute of Physics and Astronomy, University of Aarhus, 8000 Aarhus C, Denmark\\
   \email{fgj,pen,hrj@phys.au.dk}
             }

   \date{}

   \abstract{
Recent theoretical yields and chemical evolution models demonstrate that
intermediate-mass AGB stars cannot reproduce the observed abundance
distributions of O, Na, Mg, and Al. As a further observational test of this
finding, we present elemental abundance ratios [X/Fe] for 20 elements in 38
bright giants of the globular cluster NGC 6752 
based on high-resolution, high signal-to-noise spectra obtained with UVES on the VLT.
This is the most complete
spectroscopic analysis of this cluster in terms of the number of elements
considered and the number of stars in the sample. The stars span more than
1000K in effective temperature and more than 3 visual magnitudes along the red
giant branch. 
None of the abundance ratios [X/Fe] show a correlation with
evolutionary status. For Si and heavier elements, the small scatter in [X/Fe]
may be attributable to the measurement uncertainties. Our mean abundance ratios
[X/Fe] are in good agreement with previous studies of this cluster and are also
consistent with other globular clusters and field stars at the same metallicity.
The mean abundance ratios [Ba/Eu] and [La/Eu] exhibit values, in agreement with
field stars at the same metallicity, that lie approximately midway between the
pure $r$-process and the solar ($s$-process + $r$-process) mix, indicating that
AGB stars have played a role in the chemical evolution of the proto-cluster
gas. 

For the first time, we find possible evidence for an abundance variation
for elements heavier than Al in this cluster. We find a correlation between
[Si/Fe] and [Al/Fe] which is consistent with the abundance anomalies being
synthesized via proton captures at high temperatures. Leakage from the Mg-Al
chain into $^{28}$Si may explain the Si excess in stars with the highest
[Al/Fe]. We identify correlations between [Y/Fe] and [Al/Fe], [Zr/Fe] and
[Al/Fe], and [Ba/Fe] and [Al/Fe] suggesting that Y, Zr, and Ba abundances may
increase by about 0.1 dex as Al increases by about 1.3 dex. While the
correlations are statistically significant, the amplitudes of the variations
are small.  If the small variations in Y, Zr, and Ba are indeed real, then the
synthesis of the Al anomalies must have taken place within an unknown class of
stars that also ran the $s$-process. 
   }

   \maketitle
%

\section{Introduction}

\label{sec:intro}

Galactic globular clusters have provided excellent opportunities to refine
our understanding of stellar structure, stellar evolution, the formation of
the Milky Way, and the age of the universe.  While globular clusters 
constitute only 2\% of the mass of the halo \citep{freeman02}, they
are important targets to study because they are the oldest Galactic objects 
for which reliable ages have been obtained (\citealt*{vsb96};\citealt{gratton03c}). 
\citet{peebles68} suggested that globular clusters were the first bound 
systems to have formed in the protogalactic era and recently \citet{west04} 
suggested that by studying extragalactic globular clusters, the formation 
history of galaxies can be reconstructed.  However, it is important 
to recognize that our understanding of the origin and evolution of the closest 
Galactic 
globular clusters is far from complete.  Specifically, there is still no 
satisfactory explanation for the star-to-star abundance variations of light 
elements that is found in every well observed cluster.

Spectroscopic observations allow us to measure the chemical compositions of 
individual cluster stars which can provide clues to the formation 
and evolution of globular clusters (e.g., see review by
\citealt*{gratton04}).  \citet*{helfer59} presented 
the first comprehensive abundance analysis of globular cluster stars.  The first
systematic study of composition differences between various globular clusters 
was carried out by \citet{cohen78,cohen79,cohen80,cohen81}.  \citet{popper47}
first discovered a CN strong giant in M 13 and variations  
of CN were later seen among giants in M 5 and M 10 by \citet{osborn71}. 
Since the identification of large 
Na variations in giants of M 13 by \citet{peterson80}, many spectroscopic
analyses of individual cluster stars have focused upon the origin of the 
star-to-star abundance variations of light elements 
(e.g., \citealt{kraft94,kraft97,sneden97,sneden04a}).  These variations consist of 
differences in and correlations between the abundances of the 
light elements C, N, O, Na, Mg, and Al seen in every well studied Galactic
globular cluster.  Although the amplitude of the 
abundance variation may differ from cluster to cluster, there is a common 
pattern: the abundances of C and O are low when N is high and O and Na are 
anticorrelated as are Mg and Al.  

Most abundance analyses using high resolution spectra of cluster stars 
have been performed upon giants.  The advent of 8m class telescopes with
efficient high resolution echelle spectrographs has allowed 
observers to reach down 
to main sequence turn-off stars and early subgiants 
in the brightest clusters.  While star-to-star abundance variations of C 
and N were known to exist in main sequence stars of the globular cluster 47 Tuc 
\citep*{hesser78,hesser80,bell83}, these variations have now been found 
in other clusters (e.g., 
\citealt{cannon98,cohen99a}; \citealt*{briley02,cohen02}).  Recently, 
variations of O, Na, Mg, and Al have been observed in main sequence stars
\citep{gratton01,ramirez03,cohen05}.  

The two explanations for the abundance 
variations, the evolutionary and primordial 
scenarios, agree that the most likely mechanisms responsible for altering
the light element abundance ratios are proton-capture reactions (CNO-cycle, Ne-Na 
chain, and Mg-Al chain). 
In the evolutionary scenario, the abundance variations are due to internal 
nucleosynthesis and mixing within the observed stars.  
To effect changes to surface abundances of elements participating in the 
Ne-Na and Mg-Al chains requires extremely deep and extensive mixing to 
very hot layers. This may just be conceivable for red giants but is
definitely excluded as a possibility for main sequence stars. 
In the primordial scenario, the present stars either formed from gas
of an inhomogeneous composition with the O and related anomalies
present in pockets contaminated by ejecta from H-burning layers
of asymptotic giant branch (AGB) or other stars or accreted such 
ejecta after formation. 

The abundance ratios of heavy elements in cluster stars has attracted less 
attention.  Yet these elements 
may also offer much
insight into the nucleosynthetic history of globular clusters \citep{sneden04}.
Self-consistent analyses measuring a large number of elements in numerous stars 
within a given cluster \citep{M4,M5,ramirez02,ramirez03,cohen04,cohen05} 
are vital for 
our understanding of globular cluster chemical evolution.  
Here, we present an analysis of heavy elements in the cluster
NGC 6752 which along with M 13 exhibits the largest spread in
the light element abundances. Previous 
analyses of this cluster include \citet{dacosta80}, \citet{cottrell81}, 
\citet{norris81}, \citet{ss91}, \citet{norris95}, \citet{minniti96}, 
\citet{gratton01}, \citet{grundahl02}, \citet{6752}, \citet{james04}, 
\citet*{csp04}, \citet{james04b}, and \citet{carretta04} 
where each study focused upon a handful of 
abundance ratios and/or a small number of stars.  
In this paper, we present abundance ratios for 20 elements in 38 bright giants
of NGC 6752.  We explore the homogeneity of
the heavy element abundances as well as compare the abundances of various 
iron-peak, neutron-capture, and alpha elements with field stars and other 
globular clusters.  Such measurements will provide a more detailed insight into 
the chemical evolution of this globular cluster.

\section{Target selection, observations, and reduction}

The targets for this study were drawn from the $uvby$ photometry of 
\citet{grundahl99}.  The sample consists of 17 stars near the tip
of the red giant branch (RGB) and 21 stars near the bump of the RGB.
The observations were carried out in service mode with 
the UVES instrument \citep{uves} on the ESO VLT UT2 telescope.
The stars near the RGB tip were observed at a resolving power
R$\equiv \lambda/\Delta\lambda$=110,000 with signal-to-noise ratios 
(S/N) ranging from 250 per pixel in the cooler and brighter stars to 150 
per pixel in the warmer and fainter stars.  The stars near the RGB bump 
were observed with a resolving power R=60,000 with S/N=100 per pixel.
For the RGB bump stars, the abundances of O, Na, Mg, and Al were
presented in \citet{grundahl02} and for the RGB tip stars, the
abundances of O, Na, Mg, and Al and Mg isotope ratios were presented
in \citet{6752}.
For a complete description of the target selection, observations, and
data reduction see \citet{grundahl02} and \citet{6752}.  

Derivation of the stellar parameters was also described in 
\citet{grundahl02} and \citet{6752}.  Briefly, \teff~were derived 
from the \citet{grundahl99} $uvby$ photometry using the \citet*{alonso99b} 
\teff:[Fe/H]:color relations based on the infrared flux method.  
Surface gravities were estimated using the stellar
luminosities and derived \teff.  To estimate the luminosity we assumed 
a stellar mass of 0.84 $M_\odot$, an apparent distance modulus of
$(m-M)_V\,=\,13.30$, a reddening E$(B-V)\,=0.04$ \citep{harris96},
and bolometric corrections were taken from a 14 Gyr isochrone with 
[Fe/H]=$-$1.54 from \citet{vandenberg00}.  The microturbulence
was derived in the usual way by requiring that the abundances from
Fe\,{\sc i} lines be independent of the measured equivalent width.
The stellar parameters for the program stars are
presented in Table \ref{tab:param}.  

\input{2916.t1}

For each star, we started by measuring the abundance of Fe.  For
the adopted model parameters, a stellar atmosphere was taken from the
\citet{kurucz93} local thermodynamic equilibrium (LTE) stellar atmosphere 
grid.  We interpolated within the grid when necessary to obtain a model 
with the required \teff, log $g$, and [Fe/H].  The model was 
used with the LTE stellar line analysis program {\sc Moog} 
\citep{moog}.  The equivalent width (EW) of a line was measured using
routines in IRAF\footnote{IRAF is distributed by the National Optical Astronomy 
Observatories, which are operated by the Association of Universities for 
Research in Astronomy, Inc., under cooperative agreement with the National 
Science Foundation.} where in general a Gaussian profile was fitted to an
observed profile.  Assuming a solar metallicity of 
$\log\epsilon$(Fe) = 7.50, we obtain 
[Fe/H] = $-1.61 (\sigma=0.02)$ for NGC 6752 after excluding the 
star NGC6752-7 (B2438) due to its discrepant iron abundance. 
(This outlier is most likely the result of a photometric blend 
which affected the temperature and gravity estimates. 
In a $v-y$ versus $V$ diagram (Figure 1 in \citealt{6752}, it
lies to the blue of the RGB.) 
Despite the stars spanning a large range on the RGB (10.7 $\le$ V $\le$ 14.2,
3900 $\le$ \teff(K) $\le$ 4900, 0.3 $\le$ log $g$ $\le$ 2.4), we find that
the iron abundance is constant from star-to-star. 
(The $gf$ values for Fe\,{\sc i} and Fe\,{\sc ii} were presented in
\citet{6752} from which we derived $\log\epsilon$(Fe)$_\odot$ = 7.50
using a Kurucz model.) 

Previous measurements of the Fe abundance include [Fe/H]=$-$1.54 \citep{zinn84},
$-$1.58 \citep{minniti93}, $-$1.52 \citep{norris95}, $-$1.42 
\citep{carretta97}, $-$1.42 \citep{gratton01}, and $-$1.62 
\citep{grundahl02}.  \citet{kraft03} found
[Fe/H]$_{\rm I}=-1.50$ and [Fe/H]$_{\rm II}=-1.42$ using
Kurucz models and [Fe/H]$_{\rm I}=-1.51$ and [Fe/H]$_{\rm II}=-1.50$
using MARCS models.  Recently, \citet{james04} measured [Fe/H]=$-1.49 
(\sigma=0.07)$ and \citet{csp04} derived [Fe/H]=$-1.58 (\sigma=0.16)$.
The various studies employed a different set of lines and $gf$ values
in their analysis of giants or unevolved stars. 
While our Fe abundance is slightly lower than other investigators, we conclude 
that all studies are in reasonable agreement within the uncertainties in the 
stellar parameters. 

An alternative method to derive \teff~is by insisting that the abundance
from Fe lines be independent of the lower excitation potential, that is,
excitation equilibrium.  We note that 
our adopted \teff~based on photometry satisfies excitation equilibrium. 
To derive the surface gravity, an alternative method is to force the
abundance from neutral Fe lines to equal the abundance from singly
ionized Fe lines, that is, ionization equilibrium.  For our adopted surface
gravities, we note that ionization equilibrium is satisfied for all but the 
three coolest stars.  In these coolest stars, the abundance from Fe\,{\sc i} 
lines was in agreement with warmer stars.  However, 
the abundance from Fe\,{\sc ii} lines appeared to increase slightly in 
these coolest stars where the maximum discrepancy 
was Fe\,{\sc ii} $-$ Fe\,{\sc i} = 0.2 dex in the coolest star. We suggested 
that a mild revision of the temperature scale would ensure that all stars 
gave the same Fe abundance from neutral and ionized lines \citep{6752}.  
This would be achieved by a temperature correction running from 
an increase of \teff~by 100K at 3900K and vanishing at about 4200K.

\section{Elemental abundances}

\begin{figure}[ht!]
\centering
\includegraphics[width=8.0cm]{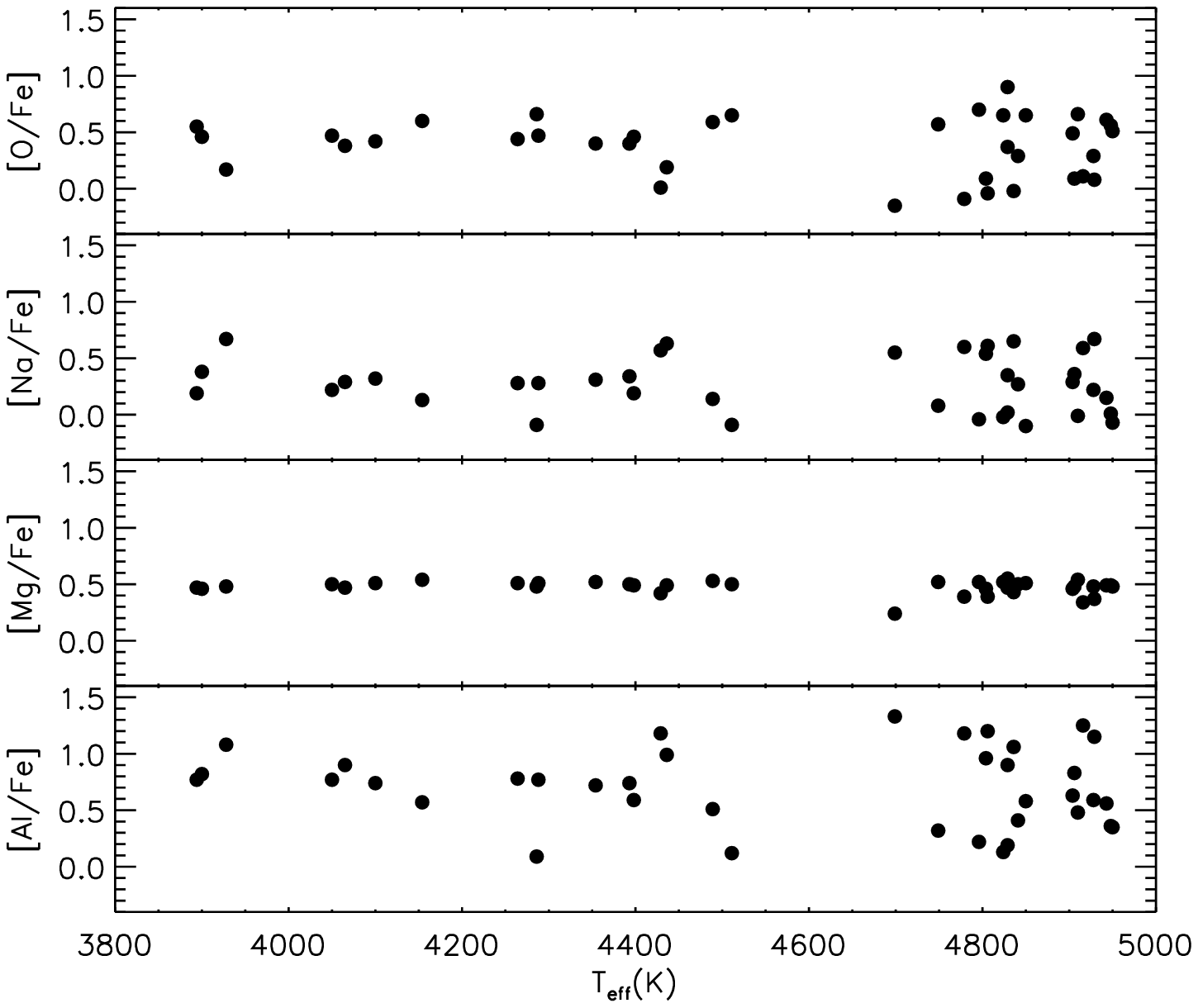}
\caption{[X/Fe] ratios for O, Na, Mg, and Al versus \teff.
\label{fig:ONaMgAl}}
\end{figure}

\begin{figure}[ht!]
\centering
\includegraphics[width=8.0cm]{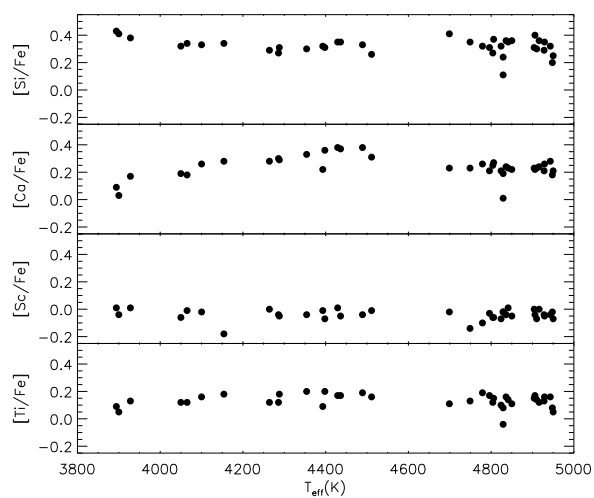}
\caption{[X/Fe] ratios for Si, Ca, Sc, and Ti versus \teff.
\label{fig:SiCaScTi}}
\end{figure}

\begin{figure}[ht!]
\centering
\includegraphics[width=8.0cm]{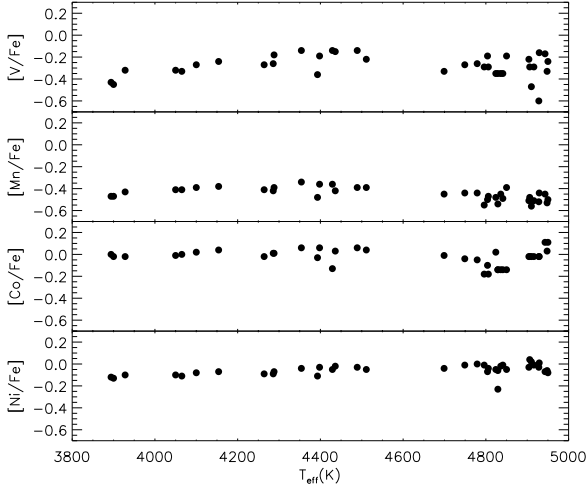}
\caption{[X/Fe] ratios for V, Mn, Co, and Ni versus \teff.
\label{fig:VMnCoNi}}
\end{figure}

\begin{figure}[ht!]
\centering
\includegraphics[width=8.0cm]{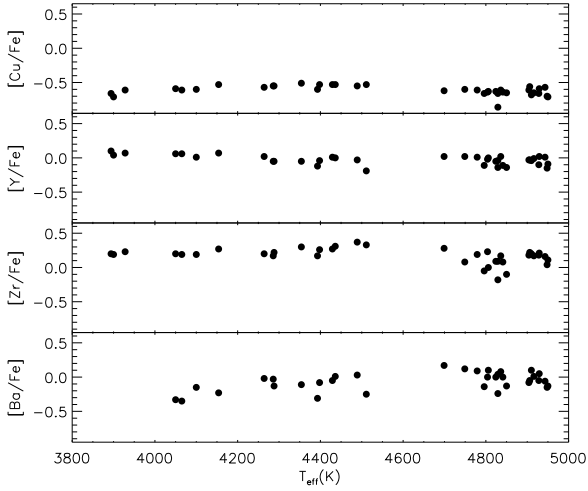}
\caption{[X/Fe] ratios for Cu, Y, Zr, and Ba versus \teff.
\label{fig:CuYZrBa}}
\end{figure}

\begin{figure}[ht!]
\centering
\includegraphics[width=8.0cm]{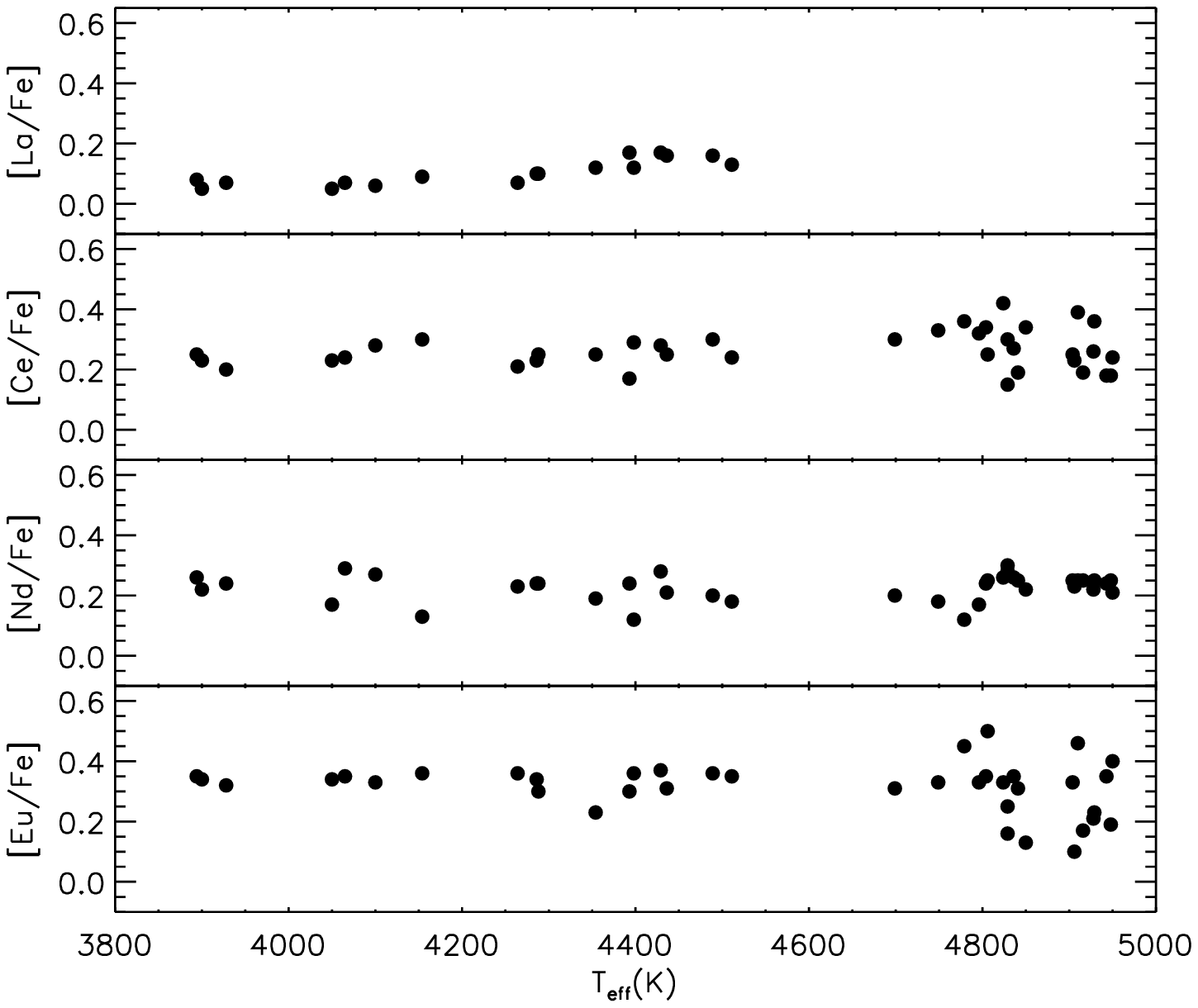}
\caption{[X/Fe] ratios for La, Ce, Nd, and Eu versus \teff.
\label{fig:LaCeNdEu}}
\end{figure}

Using the derived stellar parameters, we determined the elemental abundances
by measuring the equivalent widths of atomic lines again using routines in
IRAF.  Abundances were measured for O, Na, Mg, Al, Si, Ca, Sc, Ti, V, Mn, Co,
Ni, Cu, Y, Zr, Ba, La, Ce, Nd, and Eu.  
In general, Gaussian profiles were fitted to the observed profiles.  
For strong lines (EWs$>$80m\AA) and lines known to be affected by hyperfine 
and isotopic splitting, direct integration was used to measure the EWs.  
The elemental abundance analysis was conducted using {\sc Moog}.  We
used the Van der Waals line damping parameter 
(Uns{\" o}ld approximation multiplied by a factor recommended by the
Blackwell group). The 
line lists were compiled from \citet{kurucz95}, \citet{M5}, \citet{ramirez02}, 
and \citet{bdp03} and are presented in Table \ref{tab:line1}. Lines of
O, Na, Mg, Al, and Fe were presented in \cite{6752}. For the lines 
of Sc, V, Mn, Co, and Ba which are affected by hyperfine and/or 
isotopic splitting, we employed line lists from \citet{prochaska00}.
For lines of Cu and Eu which are affected by hyperfine and isotopic splitting, 
we used line lists from \citet{simmerer03} and \citet{eu} and assumed a solar
isotope ratio.  For La and 
Nd, we made use of the updated transition probabilities measured by 
\citet{la} and \citet{nd}.  For each of the program stars, the 
elemental abundance ratios [X/Fe] are presented in Tables \ref{tab:ab} and 
\ref{tab:ab2}.  For a particular element in a given star, the abundances 
derived from different lines are in very good agreement (in general, 
$\sigma < 0.1$ dex).
The adopted Fe abundance was the mean of all Fe lines.  Recall
that the Fe abundance from neutral and singly ionized lines agree for all but 
the three coolest stars.  Therefore, our abundance ratios are not 
affected by our choice of Fe abundance such that [X/Fe\,{\sc i}]
$\simeq$ [X/Fe\,{\sc ii}] $\simeq$ [X/Fe].
In Figures \ref{fig:ONaMgAl} to \ref{fig:LaCeNdEu}, we plot
the abundance ratios [X/Fe] versus \teff.  
The warmest stars show a
larger dispersion in [X/Fe] than the coolest stars.  This is most evident for 
V, Ce, and Eu.  We suspect that the increased spread is due 
to the lower quality of the data and the weakness of the lines.  In these 
warmer and fainter stars, lines of V, Ce, and Eu have EWs$<$
12m\AA~and the S/N is lower than for the bright RGB tip stars.  We suggest 
that the increased dispersion in the abundances of fainter 
stars is not a real feature.

In Table \ref{tab:error}, we present 
the abundance dependences on the model parameters. 
Our adopted errors are \teff~$\pm$ 30, log $g~\pm$ 0.1, and 
$\xi_t~\pm$ 0.1. Note that these are internal errors and 
underestimate the absolute errors. For \teff, we estimated the internal 
error in the following way. A polynomial fit was made to the RGB in the 
$b-y$ versus $V$ plane. The formal scatter around the relation was 
0.009mag in $b-y$ corresponding to an error of about 30K. For the
surface gravity, we assumed the basic cluster parameters for the
stars, i.e., reddening, distance, etc. Therefore, internal errors in log $g$ 
are due to the 0.01mag uncertainties in the $V$ magnitudes. This translates
into errors of the order 0.01. However, we adopted an uncertainty
of 0.1 in log $g$ since this was the minimum value that would produce 
non-zero changes in [X/Fe] for all elements. For the microturbulence, 
we plotted $\xi_t$ versus log $g$ and fitted a straight line to the data.
The scatter around the line was 0.1 km s$^{-1}$.

Within the abundance uncertainties,
none of the elemental abundance ratios show a strong dependence upon \teff,
where we take \teff~as a surrogate for evolutionary status.  
Note that our sample spans more than 1000K in \teff~and more than 3 magnitudes 
in V along the RGB including stars below the RGB bump.  

\input{2916.t2}
\input{2916.t3}
\input{2916.t4}

By comparing the predicted scatter in [X/Fe] (due to errors in the stellar 
parameters) and the measured scatter, we can understand to what extent are the 
abundance ratios constant.  We estimate the predicted scatter as 
$\sigma^2_{\rm predicted}$([X/Fe]) = $\Delta$([X/Fe]:\teff)$^2$
+ $\Delta$([X/Fe]:log $g$)$^2$ + $\Delta$([X/Fe]:$\xi_t)^2$ where 
$\Delta$([X/Fe]:\teff),  $\Delta$([X/Fe]:log $g$), and 
$\Delta$([X/Fe]:$\xi_t)$ are taken from Table \ref{tab:error} and 
represent the uncertainty in [X/Fe] due to changes in the adopted effective
temperature, surface gravity, and microturbulence.  We have not included the 
uncertainties in [X/Fe] due to errors in EWs or errors in the input 
abundance where consideration of these quantities 
would increase the predicted scatter.
Our uncertainties also do not take into account covariance terms which are 
discussed by \citet{mcwilliam95} and \citet{johnson02}. While 
\citet{johnson02} shows that these additional covariance terms are small,  
their contribution depends upon the strength of the lines, the line lists, and
the method for determining stellar parameters.  Nevertheless, adding more terms such 
as errors in the EWs or input abundance would further increase our predicted 
error which already is well matched to the observed scatter.  
In Table \ref{tab:pred},
we compare the predicted and observed scatter in abundance ratios 
(star NGC6752-7 has been excluded in calculating the observed scatter).  For
O, Na, and Al we find that the predicted scatter is significantly lower than
the observed scatter, as expected given that these elements
show star-to-star abundance variations of around 1.0 dex. For Mg, the measured
scatter slightly exceeds the predicted scatter. 
While it may be tempting to conclude that
Mg shows no variation, we emphasize that the Mg 
abundance is anticorrelated with Al (see \citealt{grundahl02} and \citealt{6752}) 
and therefore we argue that Mg shows a
small star-to-star variation despite the fair agreement between the predicted 
and observed scatter.  For elements
heavier than Al, the predicted and observed scatter are well matched. 

\input{2916.t5}
\input{2916.t6}

\begin{figure*}[ht!]
\centering
\includegraphics[width=14.5cm]{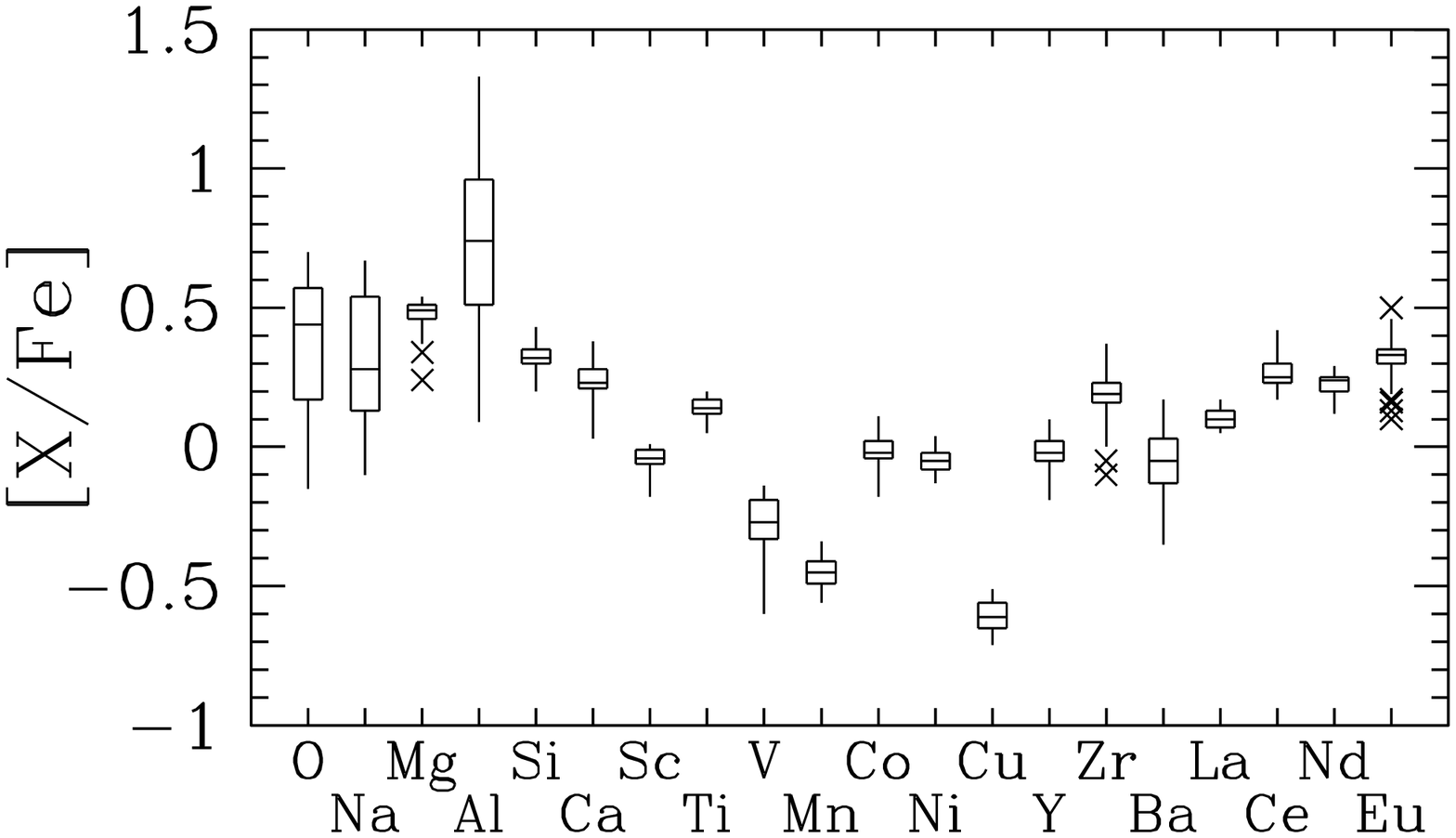}
\caption{Summary of abundance ratios.  For each element, the box represents
the interquartile range while the horizontal line is the median value.
The vertical lines extending from the box indicate the total range of the
abundance, excluding outliers.  Outliers are stars that lie more than 3 times
the interquartile range from the median and are plotted as crosses.
(Star NGC6752-7 has been omitted due to its deviating [Fe/H].) 
\label{fig:box}}
\end{figure*}

In Figure \ref{fig:box}, we plot the summary of our derived abundance ratios 
for bright giants in NGC 6752 (star NGC6752-7 has been excluded).  
Following \citet{M4,M5} and \citet{ramirez02,ramirez03},
for each element we plot a box whose upper and lower limits signify the
interquartile range (the middle 50\% of the data).  The 
line within each box identifies the median.  The vertical lines
extending from each box represent the total range of the abundance excluding
outliers.  Outliers are defined as stars that lie more than 3 times the 
interquartile range from the median.  This Figure clearly shows the large range 
of the star-to-star variation of O, Na, and Al.  As a comparison, Sc, Ti, La,  
and Nd all exhibit a very small range of abundances.    

The lack of a significant trend in abundance ratios over the large
range in \teff~may be regarded as a surprising success of an LTE analysis
employing one-dimensional plane parallel model atmospheres. Our results would 
also suggest that departures from LTE do not greatly affect the elements 
considered.

In Table \ref{tab:comp} we compare our mean abundances for elements heavier 
than Al with the \citet{norris95}, \citet{james04}, and \citet{csp04} values.  
For most elements, there is a reasonable agreement between the various studies.  
The small differences may be entirely attributed to the 
systematic offsets in stellar parameters and perhaps atomic data.  For example,
\citet{james04} derive \teff~from fitting the wings of H$_\alpha$ lines whereas
we use the \citet{alonso99b} \teff:[Fe/H]:color relations.
Certainly, the disagreement between our values and \citet{norris95} for Nd and 
La is likely due to the different atomic data where we utilize the recently
updated transition probabilities measured by 
\citet{nd} and \citet{la}.   It is probably 
no coincidence that the observed scatter in [X/Fe] for Nd and La is very low.  
Our Eu abundances differ considerably compared with \citet{norris95} and
again we identify the different transition probabilities as the likely source
of the discrepancy. 

\input{2916.t7}

\section{Comparison between globular clusters and field stars}

The behavior of O, Na, Mg, and Al in NGC 6752 (and numerous other clusters)
has been discussed extensively within the literature and is not the main focus 
of this study.  Rather, we turn our attention to elements heavier than Al.  
Following \citet{sneden04}, we compare the abundance ratios between field stars
and different globular clusters.  For the comparison field stars, we selected
the \citet{fulbright00} survey (primarily metal-poor dwarfs) and 
\citet{bdp03} survey (thin disk dwarfs) since they concentrated 
on a large number of elements in samples that exceeded 170 stars.
We also included data from the smaller samples of field dwarfs and giants 
presented by \citet{zhao90}, \citet{gratton91}, \citet{burris00}, 
and \citet{johnson02}.  For the different globular clusters, 
we selected those for which high resolution spectra had been 
obtained of large numbers of stars.  While our list of clusters is not as
extensive as \citet{sneden04}, we do include all clusters with more than 10
stars analyzed.  In Table \ref{tab:ref}, we list the clusters included for
comparisons in
the following discussion along with their metallicities placed on a uniform
scale by \citet{kraft03,kraft04}.    

\input{2916.t8}

\subsection{Alpha elements}

\begin{figure}
\centering
\includegraphics[width=8.0cm]{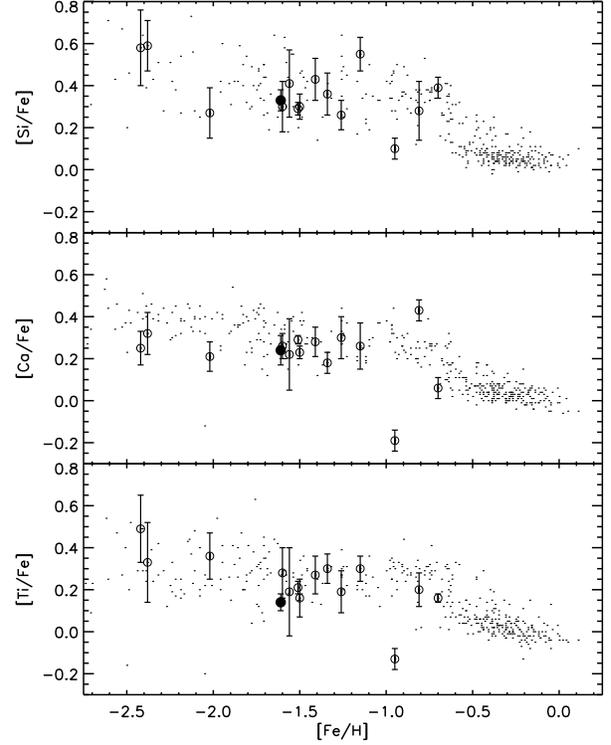}
\caption{[X/Fe] ratios for Si, Ca, and Ti versus [Fe/H].  The small dots
represent individual field stars taken from various sources (see text), the 
open circles are the mean value for different globular clusters, and the closed 
circle is mean value from this study.  The error bars on the globular clusters
represent the standard deviation.  
\label{fig:refa}}
\end{figure}

In Figure \ref{fig:refa}, we plot the mean abundances for [Si/Fe], [Ca/Fe],
and [Ti/Fe] versus [Fe/H]. We include field stars as well as mean values
for other globular clusters. 
These alpha elements are synthesized primarily in massive stars that die
as Type II supernovae whereas Fe may be produced in both Type Ia 
and Type II supernovae. The progenitors of Type II supernovae 
(i.e., massive stars) have much shorter lifetimes than the progenitors 
of Type Ia supernovae.   
A consequence of the different lifetimes and yields
is that below [Fe/H]=$-$1.0, the alpha elements are overabundant
with respect to Fe, [$\alpha$/Fe]$\simeq$0.4.  Above [Fe/H]=$-$1.0, the
abundance of the alpha elements decreases from [$\alpha$/Fe]$\simeq$0.4 to
[$\alpha$/Fe]=0.0 by solar metallicity in thin disk stars since a growing
fraction of the Fe is produced in Type Ia supernovae.  In this Figure the 
well established behavior of the alpha elements with metallicity in field 
stars is shown.  For NGC 6752, the alpha elements are also overabundant
with respect to Fe with no evidence from [$\alpha$/Fe] that Type Ia 
supernovae have contributed.  For the globular clusters, the abundances of 
Si, Ca, and Ti mimic the trends seen in the field stars.    

NGC 6752, with respect to [$\alpha$/Fe], is typical of the globular clusters,
but not all clusters follow field stars in [$\alpha$/Fe]. A notable exception
is Palomar 12. This cluster has a 
low value of [$\alpha$/Fe] compared to field stars and other globular clusters 
at the same metallicity \citep*{brown97,cohen04}.  
This cluster appears to lie in a stream extending from the Sagittarius dwarf 
spheroidal galaxy.  \citet{cohen04} showed that the abundance ratios
in Pal 12 are in good agreement with stars in the Sagittarius dwarf 
spheroidal galaxy (\citealt{bonifacio00}; \citealt*{mcwilliam03}; 
\citealt{bonifacio04,shetrone04}).  
That this cluster also appears to be younger than other Galactic globular
clusters \citep{gratton88,stetson89} reinforces the idea that Pal 12 was 
originally a globular cluster associated with the Sagittarius galaxy.  
The low [$\alpha$/Fe] may be attributed to a region in which the chemical
evolution was slow. That is, star formation
continued for a sufficiently long period such that Type Ia supernovae had time
to evolve and contaminate the stars' natal gas with iron and other products. 

\subsection{Iron-peak elements}

\begin{figure}
\centering
\includegraphics[width=8.0cm]{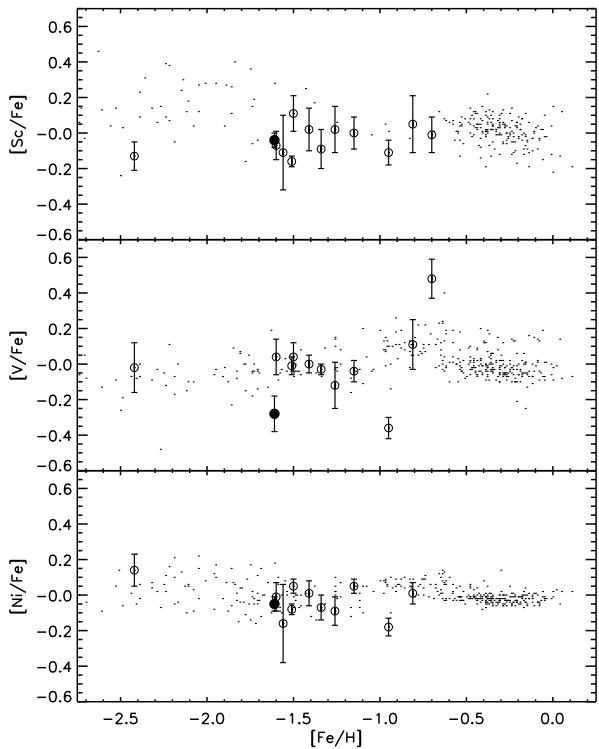}
\caption{Same as Figure \ref{fig:refa} but for Sc, V, and Ni.
\label{fig:refb}}
\end{figure}

In Figure \ref{fig:refb}, we plot the mean abundance of [Sc/Fe], [V/Fe],
and [Ni/Fe] versus metallicity in globular clusters and field stars.  In field
stars, these Fe-peak elements follow the abundance of Fe.  Again we find that 
the globular clusters seem to mimic the trends seen in the field stars.  That 
is, at a given metallicity, [X/Fe]$\simeq$0.  For Sc and Ni, NGC 6752 does not 
appear unusual when compared with other clusters: we note that our V abundance
appears low compared to field stars and other clusters.  Our mean 
value [V/Fe]=$-0.28~(\sigma=0.10)$ is lower than the \citet{norris95} 
value of [V/Fe]=$-0.01~(\sigma=0.08)$.  Our predicted scatter for V is 
well matched to the observed scatter.  
However, in a given star, the abundances from 
different V lines showed a higher scatter than for all other elements
($\sigma \simeq 0.17$ dex).  We suggest that our mean error for V 
(based on the standard deviation of [V/Fe] in all stars) is underestimated.
Pal 12 appears to be slightly underabundant
in V and Sc with respect to field stars at the same metallicity.  
As explained in the previous paragraph, \citet{cohen04} suggested that the 
peculiar abundance ratios may be evidence that Pal 12 was originally a 
globular cluster belonging to the Sagittarius galaxy.

Mn and Cu are odd-Z iron-peak elements that show subsolar abundance ratios
with respect to Fe in metal-poor field stars. 
Inspection of Figure 3 in \citet{gratton04}
shows that our mean ratios for NGC 6752 [Mn/Fe]=$-$0.45 and [Cu/Fe]=$-$0.61
are in very good agreement with other globular clusters and field stars at the
metallicity of NGC 6752 [Fe/H]=$-$1.61. 

\subsection{Neutron-capture elements}

Heavy elements are synthesized via neutron-capture through either the $s$- or
the $r$-process. The latter's site is generally identified with Type II
supernovae (i.e., the death of massive stars), and the former with AGB
stars. In general, both processes may contribute to the synthesis of a 
particular element. Dissection of the solar system abundances shows that Eu
is primarily a $r$-process product: \citet{burris00} put the $r$-process
fraction at 97\%. On the other hand, Ba in the solar mix is largely a
$s$-process product: 85\% due to that process and 15\% to the $r$-process.
Of the other elements we have measured, Y, Zr, La, and Ce approximately follow
Ba with $s$-process contributions ranging from 72\% to 81\%. 
The remaining element Nd
for the solar mix is roughly equally attributed to both processes. There is 
as yet no firm theoretical guidance on how these $r$- and $s$-process fractions
may vary with metallicity and other variables influencing the chemical
evolution of a globular cluster. 

\begin{figure}
\centering
\includegraphics[width=8.0cm]{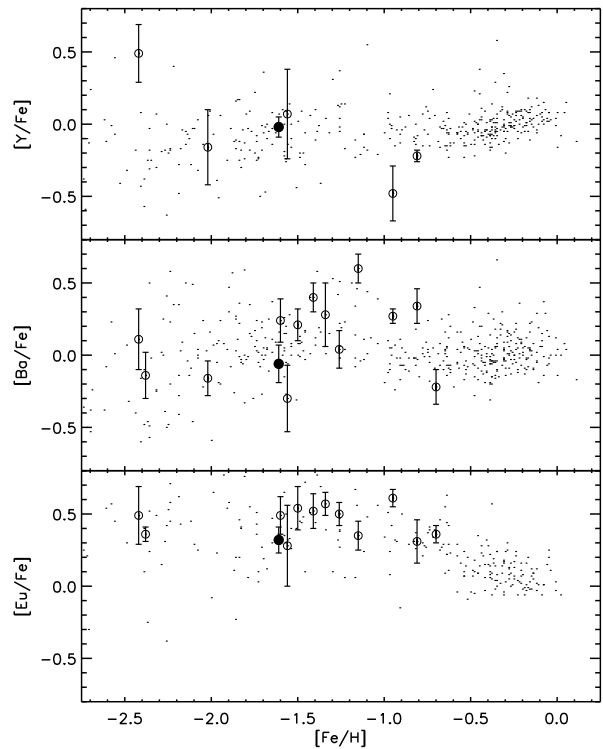}
\caption{Same as Figure \ref{fig:refa} but for Y, Ba, and Eu.
\label{fig:refc}}
\end{figure}

In Figure \ref{fig:refc}, we plot the mean abundance of [Y/Fe], [Ba/Fe], and
[Eu/Fe] versus [Fe/H] in globular clusters and field stars. The [Eu/Fe] of
the field stars rises to [Eu/Fe] $\simeq +0.5$ with little star-to-star scatter
until [Fe/H] $\simeq -2.0$. That this behavior 
mimics that of an $\alpha$-element
suggests that Eu and the $\alpha$-elements have a common origin (i.e.,
Type II supernovae) and that the relative yields from stellar nucleosynthesis
of $\alpha$-elements like Mg and Si to Eu are insensitive to the initial 
metallicity of the stars. The mean [Eu/Fe] for NGC 6752 is similar to that
of field stars and other globular clusters of NGC 6752's [Fe/H]. 
For [Y/Fe] and [Ba/Fe], the field stars do not show any obvious behavior
with metallicity. While the scatter is large, the globular cluster abundances
track the field stars. 

Some halo stars with large enhancements of neutron-capture elements 
show scaled solar $r$-process abundance distributions for Ba and heavier
elements (e.g., \citealt{cowan99,westin00,hill02,22892.03}).  
A few elements permit measurements of the isotope ratios.  In a handful of
metal-poor stars, the isotope ratios for Ba \citep{lambert02} and 
Eu \citep{sneden02,aoki03} are consistent with a scaled solar pure 
$r$-process mix.  That some metal-poor stars show 
scaled solar $r$-process abundances is evidence that the $r$-process may be 
universal, at least for heavy elements, $Z \ge 56$.  That is, whatever mechanism 
is responsible for the synthesis of the $r$-process elements (Ba and heavier
elements) in the sun may have operated at all metallicities.    

In addition to the usual dispersion in 
O, Na, Mg, and Al, \citet{sneden97} found star-to-star variations of Ba and Eu
in the globular cluster M 15 with the mean value of [Ba/Fe] and [Eu/Fe] agreeing 
well with other clusters.     
\citet{sneden04} note that while the abundances of Ba and Eu vary in M 15,
for all stars the ratio [Ba/Eu] is constant. \citet{sneden00b} re-observed
3 giants in M 15 in order to conduct a more detailed abundance analysis of 
heavy elements.  They found that the abundance ratios for Ba to Dy matched
the scaled solar-system $r$-process distribution.  
This suggests that the evolution of the heavy element abundances of M 15 is
dominated solely by explosive nucleosynthesis in 
high-mass stars with effectively no contribution from 
AGB stars. 

\citet{james04} measured [Ba/Eu]=$-0.18\pm0.11$ in 
unevolved stars in NGC 6752.  This value lies between the pure $r$-process
and the solar $s+r$ mix.  
Our [Ba/Eu]=$-0.37\pm0.16$ is slightly
lower than the \citet{james04} value, but within the errors these ratios
are in agreement.  Our value of [Ba/Eu] also lies between the pure $r$-process
and the solar $s+r$ mix and agrees with field stars at the same 
metallicity.  \citet{simmerer04} measured the ratio [La/Eu] in a large 
sample of field stars. [La/Eu] offers an alternative to [Ba/Eu] 
as a measure of the evolution of the $s$- and 
$r$-process.  La has numerous lines at visible wavelengths and therefore
provides a more reliable measure of the $s$-process than Ba whose few lines are 
often saturated.  Our value of [La/Eu]=$-0.23\pm0.10$ in bright
giants of NGC 6752 is roughly halfway between the pure $r$-process
and the solar $s+r$ mix.  This value is also in good agreement with field stars
at the same metallicity \citep{simmerer04}.  The $s$-process 
is believed to occur in low-mass to intermediate-mass AGB stars.  The solar 
$s+r$ mix therefore represents a combination of products of Type II
supernovae and AGB stars.  
That our measured ratios of [Ba/Eu] and [La/Eu] lie between the pure $r$-process
and the solar $s+r$ mix indicates that the material from which the cluster
stars formed was enriched by AGB ejecta. While these elements may have been 
synthesized in the cluster, or proto-cluster, an alternative possibility is that
they were produced in the halo prior to the formation of the cluster. The latter
option may explain the similarity in [Ba/Eu] and [La/Eu] abundances between 
field and cluster stars of the same metallicity. 
Unlike the heavier elements, for elements lighter than Ba the abundance ratios
observed in metal-poor stars do not match the pure $r$-process solar mix.  In 
representative field stars, the ratios [Sr/Ba] and [Y/Ba] show a large dispersion 
at low [Fe/H] despite having a constant [Ba/Eu] close to the pure $r$-process 
value \citep{mcwilliam98}.  These measurements (constant [Ba/Eu] combined with 
variations in [Sr/Ba] and [Y/Ba]) have led to the suggestion that there are
multiple sites for the $r$-process \citep{wasserburg96,sneden00c,qian01}.  
Our cluster value [Y/Ba]=0.03 $(\sigma=0.13)$ lies well
above the pure $r$-process value and is in good agreement with field stars
at the same metallicity.  

\section{Globular cluster chemical evolution}

\subsection{Abundance anomalies for light elements}

Having discussed elements heavier than Al, we offer some 
comments relating to the star-to-star abundance variation of
light elements.  
At the heart of the proposed
evolutionary scenarios are various mixing mechanisms required to transport
material from deep layers within the star to the outer envelope.  These mixing 
processes cannot operate in stars below the RGB bump due to 
the sharp composition discontinuity left by the deepest penetration of the 
convective envelope \citep{sweigart79}. \citet{sweigart79} also 
showed that once the composition barrier is removed by the outward 
moving H-burning shell, mixing mechanisms can tap deep layers where H-burning 
occurs through the CNO-cycles, and possibly the Ne-Na and Mg-Al chains.  Over a 
wide range
of metallicities, \citet{zoccali99} have shown that the observed luminosity
of the RGB bump agrees with theoretical predictions.  
The crucial discovery of the O-Na and Mg-Al 
anticorrelation in unevolved stars in NGC 6752 \citep{gratton01} 
was proof that the abundance variations of O, Na, Mg, and Al must be 
primarily due to a primordial scenario.
Furthermore, \citet{grundahl02} measured abundances
of O, Na, Mg, Al, and Li in NGC 6752 giants. They 
showed that Li is present in stars below the RGB bump, but absent in stars
above the bump.  That is, mixing in cluster stars cannot occur below the bump
but above the bump Li is destroyed by mixing.  However, \citet{grundahl02} 
also demonstrated that the anticorrelations between O-Na and Mg-Al are
present below the bump with no obvious change in amplitude or in mean
abundance, reinforcing 
the evidence provided by \citet{gratton01} that the O, Na, Mg, and Al variations
must have a primordial origin. Note that an ``evolutionary'' component is
essential to explain the C, N, and Li abundances as well as C isotope ratios 
that show a dependence on evolutionary status. 

\subsection{AGB stars}

Based on overabundances
of Na and Al in CN strong stars of NGC 6752, \citet{cottrell81} first proposed 
a primordial scenario to explain the abundance variations 
in which intermediate-mass AGB stars pollute the 
proto-cluster gas.  The envelopes of metal-poor intermediate-mass AGB 
stars qualitatively have the correct
composition to produce the observed abundance anomalies. Two possibilities
exist for the AGB pollution scenario. In the first scenario, AGB stars
pollute the proto-cluster gas from which the present cluster members form
as suggested by \citet{cottrell81}. 
In the second scenario, the present cluster members accrete material 
ejected from AGB stars. Low-mass main sequence stars have thin convection zones
whereas evolved giants have deep convective envelopes. If the abundance anomalies
were only present in the thin convective zone during the main sequence, they 
would be diluted as the star ascends the giant branch. Observations show that
this does not occur. While calculations by \citet{thoul02} have shown that 
cluster ``stars can accrete an appreciable fraction of their initial mass'', it
is perhaps more likely that the stars were born with inhomogeneous compositions. 

If the abundance variations are due to differing degrees of pollution of natal 
clouds from intermediate-mass AGB stars, should there be other elements besides
O-Al that display a star-to-star variation? There are two possible effects that
we identify. The first is that H-burning at high temperatures in AGB stars will
alter the H/He ratio. The most obvious direct effect is that
the hydrogen abundance should be lower and hence we would expect a higher [Fe/H]
in stars with the highest Al abundance. There is an additional consequence
of differing H/He ratios on the
atmosphere. The increase in He and corresponding decrease in H lowers the 
continuous opacity (H$^-$) per gram. For an atmosphere with increased He, 
the appearance of the spectrum would be equivalent to an increase in heavy element
abundances and surface gravity (see \citealt{bv79} for a detailed discussion). It 
would be 
of great interest to do a detailed calculation of the expected change in [X/Fe] 
due to differing ratios of H/He. The second effect that we may expect is 
a correlation between $s$-process elements and say Al abundance since
$s$-process elements are synthesized in AGB stars.
However, in metal-poor intermediate-mass AGB stars, the O-Al abundances are
altered in the hot-bottom convective envelope. The $s$-process elements are
synthesized in thermal pulses and dredge-up. If this is a correct division of 
responsibility, it is possible that the $s$-process abundances may show 
little dependence on the O-Al abundance variations. 
However, as far as we are aware, 
theoretical predictions of $s$-process yields from metal-poor 
intermediate-mass (M$ > 3$M$_\odot$) AGB stars are rare. 
\cite{travaglio04} carried out calculations 
for a range of masses and metallicities but did not publish 
yields for each model. 

\subsection{Quantitative problems with the AGB pollution scenario}

While the AGB pollution scenario may offer an appealing qualitative explanation 
for the observed abundance anomalies, a more quantitative consideration reveals a 
number of serious problems (see \citealt{lattanzio04} for a review). The AGB models
predict much higher ratios of $^{25}$Mg/$^{24}$Mg and $^{26}$Mg/$^{24}$Mg
when O is depleted \citep{denissenkov03}. AGB models also cannot produce the
observed pattern of low C and O abundances along 
with high N abundances \citep{denissenkov04}.
A chemical evolution model of NGC 6752 constructed by \citet{fenner04} 
incorporating recent AGB yields reproduced the Na and Al dispersion. 
However O was not sufficiently depleted, Mg was produced rather than destroyed, 
C+N+O was not constant, and $^{25}$Mg should be correlated with $^{26}$Mg. 
\citet{fenner04} note that all of these problems arise from the addition
of He-burning products into the AGB ejecta and that a generation of AGB
stars that experience hot-bottom burning but no dredge-up of He-burning
products might provide a better match to the observations. 
We caution that globular cluster chemical evolution modelling is in a developing
phase and that the AGB yields are critically dependent on the treatment of
convection. While \citet{fenner04} found that Mg increases with Al, 
\citet{denissenkov04} discuss $^{24}$Mg destruction with increasing Al. 
However, the current theoretical yields and chemical evolution
models do not favor the AGB pollution scenario (with intermediate-mass AGBs) 
as the mechanism responsible
for the star-to-star abundance variations in globular clusters. 

An alternative pollution scenario has been proposed by 
\citet{denissenkov04}. In this case, the star-to-star variations
may result from mass transfer in binaries in which the more massive
star (now a white dwarf) was an RGB and/or AGB star slightly more massive
than the present main sequence turn-off cluster stars. To explain the
abundance variations, these RGB/AGB
stars must have experienced extra mixing during the course of
their evolution. That the O-Na and Mg-Al anticorrelations are not seen
in field stars would imply a fundamental difference between field 
and globular cluster stars. 

\subsection{Additional evidence to probe the origin of the abundance anomalies:
heavy element variations}

A major goal of our present study is to search for heavy element 
variations. Our very accurate abundance ratios can 
be used to search for correlations between 
heavy elements and Al. This will provide additional clues to the origin of the
abundance variations. That the observed scatter is well matched to the
predicted scatter is no guarantee that the abundances for Si and heavier 
elements are constant. For example, 
the observed and predicted dispersions for Mg are in agreement but Mg and Al 
are anticorrelated and therefore Mg shows a small star-to-star variation. Do 
any other elements show correlations with Al? 

\begin{figure}
\centering
\includegraphics[width=8.0cm]{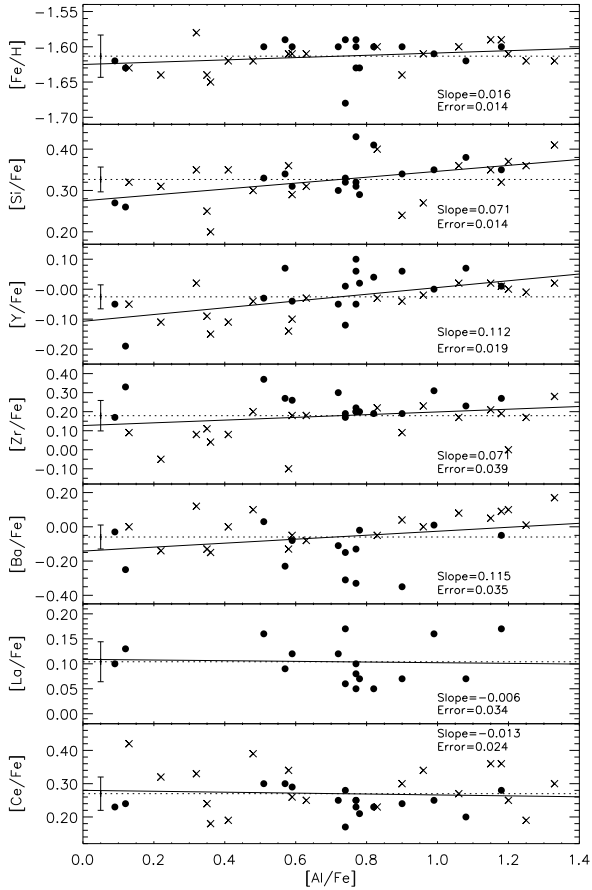}
\caption{Abundance ratios [Fe/H], [Si/Fe], and [X/Fe] (for $s$-process elements
Y, Zr, Ba, La, and Ce) versus [Al/Fe]. Crosses represent the bump stars and
filled circles represent the tip stars. The error bar shows the $\pm$1-$\sigma$
predicted error
from Table \ref{tab:pred}. The dotted line is the mean abundance and the solid
line is the linear least squares fit to the data (slope and associated error
are included). Star NGC6752-7 has been omitted due to its deviating [Fe/H].
\label{fig:al_s}}
\end{figure}

In Figure \ref{fig:al_s}, we plot [Fe/H], [Si/Fe],
and $s$-process elements (Y, Zr, Ba, La, and Ce) versus [Al/Fe]. 
Different symbols are used for the bump and tip stars which may help 
reveal whether correlations are real or merely artefacts of \teff~effects. 
There is a hint that the iron abundance [Fe/H] may increase with
increasing [Al/Fe]. If verified through a more careful differential 
analysis, this may indicate that the He abundance increases with increasing [Al/Fe]
as expected if the abundance anomalies are produced via H-burning at high
temperatures. We find a statistically significant 
correlation between [Si/Fe] and [Al/Fe]. Such 
a trend would arise if the reaction $^{27}$Al(p,$\gamma$)$^{28}$Si takes place
rather than $^{27}$Al(p,$\alpha$)$^{24}$Mg within the Mg-Al chain. 
Hot-bottom burning in intermediate-mass AGB stars can produce $^{28}$Si from 
proton capture on $^{27}$Al, though the Si yields are expected to be 
small \citep{karakas03t}. 
While the small Si excess may be due to production of $^{29}$Si and $^{30}$Si via
neutron capture in the He shell of AGB stars, the total Si abundance would not
change unless some leakage from the Mg-Al cycle into $^{28}$Si takes place. 
A statistically significant
slope is also evident between [Y/Fe] and [Al/Fe]. 
For Zr, the bump stars show a clear correlation between [Zr/Fe] and [Al/Fe]
where the Zr abundances are derived from the 5112\AA~Zr\,{\sc ii} line. The Zr 
abundances for the tip stars come from both Zr\,{\sc i} and Zr\,{\sc ii} lines. 
If we consider only abundances from Zr\,{\sc ii} lines, the slope (0.12) 
and error (0.04) for the linear least squares fit 
between [Zr/Fe] and [Al/Fe] is more significant. [Ba/Fe] is also correlated 
with [Al/Fe], the third $s$-process element that shows evidence for a variation 
with [Al/Fe]. 
On the other hand, the Eu abundance [Eu/Fe] 
appears uniform across the sample. There is no
correlation between Eu and Al abundances across the more than 1 dex spread
in the Al abundance, suggesting that the source of the Al (and O, Na, and Mg) 
abundance variations is not producing Eu.

\begin{figure}
\centering
\includegraphics[width=8.0cm]{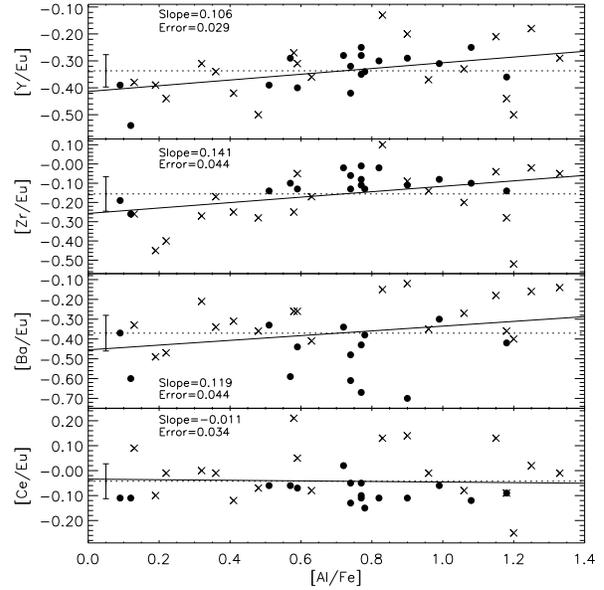}
\caption{Same as Figure \ref{fig:al_s} except for abundance ratios [Y/Eu], [Zr/Eu],
[Ba/Eu], and [Ce/Eu] versus [Al/Fe]. Here we use only the abundances from
ZrII lines. \label{fig:al_s2}}
\end{figure}

Perhaps the most direct and accurate indicator of a dispersion in heavy element 
abundances will be provided by relative abundance ratios between heavy elements. 
In Figure \ref{fig:al_s2}, we plot [Y/Eu], [Zr/Eu], [Ba/Eu], and [Ce/Eu]
versus [Al/Fe]. In particular, abundances for Y, Zr, Ce, and Eu come from 
relatively unsaturated lines of ions in all cases. Again, we find correlations
between heavy element abundance ratios and [Al/Fe]. 

While the correlations between [X/Fe] and [Al/Fe] as well as [X/Eu] and [Al/Fe]
are statistically significant, we note that the 
amplitudes of the abundance variations are small. The 1.3 dex increase in [Al/Fe]
is accompanied by roughly 0.1 dex increases in [X/Fe]. As a comparison, for 
[Mg/Fe] versus [Al/Fe] the formal slope is 0.13 with an uncertainty of 0.01. 
It is possible that errors in the stellar parameters are producing the
correlations. However, our linear least squares fits take into account the 
uncertainties due to errors in the stellar parameters. 
Since [Al/Fe] and [Na/Fe] span more than 1 dex and the Al-rich/poor stars 
are not correlated with \teff, it will be difficult for atmospheric parameter 
errors to produce a slope between heavy elements and Al. 

For the first time, we provide evidence that the $s$-process elements Y, Zr, 
and Ba are correlated with Al and may therefore exhibit a small 
star-to-star variation. \citet{james04} did not find 
correlations between Sr, Y, Ba, or Eu and Al in unevolved stars though their 
spectra were taken with a lower resolving power and S/N. 
Our results suggest that the stars in which the abundance 
anomalies were produced also synthesized $s$-process elements to
a small degree. While AGB stars
may be the most promising candidates for explaining correlations between say
Y and Al, recent AGB yields and chemical evolution models show 
that intermediate-mass AGB stars cannot explain the light element abundance 
variations. If the variations in Y, Zr, and Ba are real, then the star-to-star
dispersion for O-Al abundances 
may be due to differing degrees of pollution from stars that also ran the
$s$-process: perhaps an unknown class of AGB stars. 

The lack of significant correlations between the
other $s$-process elements (La and Ce) and [Al/Fe] may be due to measurement 
uncertainties, though both La and Ce have smaller predicted and observed 
dispersions than Zr and Ba. On the other hand, the lack of trends between La and 
Al or Ce and Al might be a real effect that can be used to probe the stars in which 
the Al anomalies were synthesized (presumably the same stars are also 
responsible for producing the C, N, O, Na, and Mg variations). 

\section{Summary and concluding remarks}

Globular clusters are the oldest Galactic objects and are considered by some to
be the basic building blocks of galaxies.  Yet, our understanding of their 
chemical evolution is incomplete.  Here we present elemental abundance ratios 
[X/Fe] for 20 elements in 38 bright giants 
in the globular cluster NGC 6752 to study the chemical evolution
of this cluster. 
Our sample size and number of elements considered makes this study the most 
comprehensive spectroscopic abundance analysis of NGC 6752 to date.
None of 
the abundance ratios [X/Fe] show a trend with \teff, that is, evolutionary 
status.  We estimated the predicted scatter due to uncertainties in the adopted 
stellar parameters (effective temperature, surface gravity, and microturbulence).  
For all elements 
heavier than Al, the small observed scatter is well matched by the predicted scatter. 
The mean abundance ratios [X/Fe] for elements heavier than Al are in 
good agreement with previous studies of bright giants and unevolved stars in this
cluster.  The mean [X/Fe] for this cluster are also consistent with other 
globular clusters as well as field stars at the same metallicity.  
For elements heavier than Al, the nucleosynthetic
processes responsible for the evolution of the elements in field stars must
also drive the proto-globular cluster gas to its present metallicity.

The abundance ratio [Ba/Eu]=$-0.37\pm0.16$ agrees with previous studies and 
lies midway between the pure $r$-process value and 
the solar ($s$-process + $r$-process) mix.  Similarly, our measured abundance
ratio [La/Eu]=$-0.23\pm0.10$ 
(which takes advantage of recently updated transition probabilities)
lies midway between the pure $r$-process and the solar $s+r$ mix and
agrees with field stars at the same metallicity 
\citep{simmerer04}.  This demonstrates that AGB stars played a role in the
chemical evolution of the proto-cluster gas.

Mg is an example where the observed and predicted dispersion are in good agreement,
but an anticorrelation with Al indicates a star-to-star variation for Mg. We compared 
abundance ratios with [Al/Fe] to search for correlations which would reveal small
abundance variations. There was a hint of a trend 
between [Fe/H] and [Al/Fe]. If confirmed from future analyses, this would suggest
differing H/He ratios as expected if the abundance anomalies are produced from
H-burning at high temperatures. We found a correlation between [Si/Fe] and [Al/Fe]
which can be explained if the reaction 
$^{27}$Al(p,$\gamma$)$^{28}$Si is favored over $^{27}$Al(p,$\alpha$)$^{24}$Mg
at the end of the Mg-Al chain. Hot-bottom burning in intermediate-mass AGB stars
is expected to produce small amounts of Si. 
Most importantly, we found correlations between [Y/Fe] and [Al/Fe], [Zr/Fe] and 
[Al/Fe], as well as [Ba/Fe] and [Al/Fe]. These correlations offer the first
evidence for variations in $s$-process elements for NGC 6752. That these
elements are correlated with [Al/Fe] suggests that the stars responsible for
the synthesis of the Al variations (and presumably all abundance anomalies) 
also synthesized $s$-process elements. 
The stellar origins of the light and heavy element abundance variations
remain uncertain. Intermediate-mass AGB stars remain a
viable candidate as long as theoretical models contain major
uncertainties (e.g., the treatment of convection) and published
models differ over the composition of the ejecta (e.g., is
Mg enriched or depleted? [\citealt{fenner04,denissenkov04}]). 

Despite measuring the abundances of 20 elements in NGC 6752, there are at least
two more elements we wish to measure which will provide strong constraints on
the possible role of AGB stars in the chemical evolution of this 
cluster. Rb and Pb are of particular interest and have not yet been measured in 
NGC 6752 nor in any other globular cluster. Measurements of Rb 
will provide further observational constraints on the AGB pollution 
scenario since the ratio Rb/Zr is sensitive to the neutron density and therefore
mass of the AGB star (\citealt{lambert95,tomkin99}; \citealt*{busso99}; 
\citealt{abia01}). 
Intermediate-mass AGB stars are predicted to have high neutron densities
and so we would expect Rb to be overabundant with respect to Y or Zr if 
such stars have contributed to the evolution of the cluster. 
Since Pb and Bi are the main products of very metal-poor AGB stars \citep{busso99},
we may expect considerable Pb enhancements. These measurements 
will provide further insight into the chemical evolution of this globular cluster.


\begin{acknowledgements}
We thank the referee, Raffaele Gratton, for helpful comments. 
DY is grateful to Bruce Carney, Inese Ivans, Amanda Karakas, John Lattanzio, 
Jennifer Simmerer, and Jocelyn Tomkin for helpful discussions 
and insights. FG gratefully acknowledges the financial support provided by the
Carlsberg Foundation and through a grant from Professor Henning 
E.\ J{\o}rgensen. Financial support from The Instrument Center for
Danish Astrophysics (IDA) during the last phase of the project is also acknowledged.
DLL and DY acknowledge support from the Robert A.\ Welch 
Foundation of Houston, Texas. 
This research has made use of the SIMBAD database,
operated at CDS, Strasbourg, France and NASA's Astrophysics Data System.
\end{acknowledgements}


\end{document}

%% file: 2916.t1.tex
\begin{table*}
\caption{Stellar parameters for program stars 
\label{tab:param}}
\begin{tabular}{llccccccc}
\hline
Name1 &
Name2 &
RA &
Dec &
V &
\teff &
log g &
$\xi_t$ &
[Fe/H]
\\
 &
 &
(2000) &
(2000) &
 &
(K) &
(cm s$^{-2}$) &
(km s$^{-1}$) &
\\
\hline
PD1 & NGC6752-mg0 & 19:10:58 & $-$59:58:07 & 10.70 & 3928 & 0.26 & 2.70 & $-$1.62 \\
B1630 & NGC6752-mg1 & 19:11:11 & $-$59:59:51 & 10.73 & 3900 & 0.24 & 2.70 & $-$1.60 \\
B3589 & NGC6752-mg2 & 19:10:32 & $-$59:57:01 & 10.94 & 3894 & 0.33 & 2.50 & $-$1.59 \\
B1416 & NGC6752-mg3 & 19:11:17 & $-$60:03:10 & 10.99 & 4050 & 0.50 & 2.20 & $-$1.60 \\
\nodata & NGC6752-mg4 & 19:10:43 & $-$59:59:54 & 11.02 & 4065 & 0.53 & 2.20 & $-$1.60 \\
PD2 & NGC6752-mg5 & 19:10:49 & $-$59:59:34 & 11.03 & 4100 & 0.56 & 2.10 & $-$1.59 \\
B2113 & NGC6752-mg6 & 19:11:03 & $-$60:01:43 & 11.22 & 4154 & 0.68 & 2.10 & $-$1.59 \\
\nodata & NGC6752-mg8 & 19:10:38 & $-$60:04:10 & 11.47 & 4250 & 0.80 & 2.00 & $-$1.68 \\
B3169 & NGC6752-mg9 & 19:10:40 & $-$59:58:14 & 11.52 & 4288 & 0.91 & 1.90 & $-$1.63 \\
B2575 & NGC6752-mg10 & 19:10:54 & $-$59:57:14 & 11.54 & 4264 & 0.90 & 1.80 & $-$1.63 \\
\nodata & NGC6752-mg12 & 19:10:58 & $-$59:57:04 & 11.59 & 4286 & 0.94 & 1.80 & $-$1.62 \\
B2196 & NGC6752-mg15 & 19:11:01 & $-$59:57:18 & 11.68 & 4354 & 1.02 & 1.90 & $-$1.60 \\
B1518 & NGC6752-mg18 & 19:11:15 & $-$60:00:29 & 11.83 & 4398 & 1.11 & 1.80 & $-$1.60 \\
B3805 & NGC6752-mg21 & 19:10:28 & $-$59:59:49 & 11.99 & 4429 & 1.20 & 1.80 & $-$1.60 \\
B2580 & NGC6752-mg22 & 19:10:54 & $-$60:02:05 & 11.99 & 4436 & 1.20 & 1.80 & $-$1.61 \\
B1285 & NGC6752-mg24 & 19:11:19 & $-$60:00:31 & 12.15 & 4511 & 1.31 & 1.90 & $-$1.63 \\
B2892 & NGC6752-mg25 & 19:10:46 & $-$59:56:22 & 12.23 & 4489 & 1.33 & 1.70 & $-$1.60 \\
\nodata & NGC6752-0 & 19:11:03 & $-$59:59:32 & 13.03 & 4699 & 1.83 & 1.47 & $-$1.62 \\
B2882 & NGC6752-1 & 19:10:47 & $-$60:00:43 & 13.27 & 4749 & 1.95 & 1.41 & $-$1.58 \\
B1635 & NGC6752-2 & 19:11:11 & $-$60:00:17 & 13.30 & 4779 & 1.98 & 1.39 & $-$1.59 \\
B2271 & NGC6752-3 & 19:11:00 & $-$59:56:40 & 13.41 & 4796 & 2.03 & 1.42 & $-$1.64 \\
B611 & NGC6752-4 & 19:11:33 & $-$60:00:02 & 13.42 & 4806 & 2.04 & 1.40 & $-$1.61 \\
B3490 & NGC6752-6 & 19:10:34 & $-$59:59:55 & 13.47 & 4804 & 2.06 & 1.40 & $-$1.61 \\
B2438 & NGC6752-7 & 19:10:57 & $-$60:00:41 & 13.53 & 4829 & 2.10 & 1.33 & $-$1.84 \\
B3103 & NGC6752-8 & 19:10:45 & $-$59:58:18 & 13.56 & 4910 & 2.15 & 1.33 & $-$1.62 \\
B3880 & NGC6752-9 & 19:10:26 & $-$59:59:05 & 13.57 & 4824 & 2.11 & 1.38 & $-$1.63 \\
B1330 & NGC6752-10 & 19:11:18 & $-$59:59:42 & 13.60 & 4836 & 2.13 & 1.37 & $-$1.60 \\
B2728 & NGC6752-11 & 19:10:50 & $-$60:02:25 & 13.62 & 4829 & 2.13 & 1.32 & $-$1.64 \\
B4216 & NGC6752-12 & 19:10:20 & $-$60:00:30 & 13.64 & 4841 & 2.15 & 1.34 & $-$1.62 \\
B2782 & NGC6752-15 & 19:10:49 & $-$60:01:55 & 13.73 & 4850 & 2.19 & 1.35 & $-$1.61 \\
B4446 & NGC6752-16 & 19:10:15 & $-$59:59:14 & 13.78 & 4906 & 2.24 & 1.32 & $-$1.60 \\
B1113 & NGC6752-19 & 19:11:23 & $-$59:59:40 & 13.96 & 4928 & 2.32 & 1.29 & $-$1.61 \\
\nodata & NGC6752-20 & 19:10:36 & $-$59:56:08 & 13.98 & 4929 & 2.33 & 1.32 & $-$1.59 \\
\nodata & NGC6752-21 & 19:11:13 & $-$60:02:30 & 14.02 & 4904 & 2.33 & 1.29 & $-$1.61 \\
B1668 & NGC6752-23 & 19:11:12 & $-$59:58:29 & 14.06 & 4916 & 2.35 & 1.27 & $-$1.62 \\
\nodata & NGC6752-24 & 19:10:44 & $-$59:59:41 & 14.06 & 4948 & 2.37 & 1.15 & $-$1.65 \\
\nodata & NGC6752-29 & 19:10:17 & $-$60:01:00 & 14.18 & 4950 & 2.42 & 1.26 & $-$1.64 \\
\nodata & NGC6752-30 & 19:10:39 & $-$59:59:47 & 14.19 & 4943 & 2.42 & 1.27 & $-$1.62 \\
\hline
\end{tabular}

Note. $-$-- PD1 and PD2 are from \citet{penny86} and the B xxxx names are from 
\citet{buonanno86}.

\end{table*}

%% file: 2916.t2.tex
\begin{table} 
\caption{Atomic line list\label{tab:line1}}
\begin{tabular}{lccrc}
\hline
Species &
Wavelength (\AA) &
EP (eV) &
log $gf$ &
Ref.$^{\mathrm{a}}$ \\
\hline
Si\,{\sc i}  & 5645.61 & 4.93 &  $-$2.14  & RC02 \\
Si\,{\sc i}  & 5665.55 & 4.92 &  $-$2.04  & RC02 \\
Si\,{\sc i}  & 5690.43 & 4.93 &  $-$1.87  & RC02 \\
Si\,{\sc i}  & 5701.11 & 4.93 &  $-$2.05  & RC02 \\
Si\,{\sc i}  & 5948.55 & 5.08 &  $-$1.23  & RC02 \\
Si\,{\sc i}  & 6142.49 & 5.62 &  $-$1.48  & IK01 \\
Si\,{\sc i}  & 6145.02 & 5.61 &  $-$1.44  & RC02 \\
Si\,{\sc i}  & 6155.13 & 5.62 &  $-$0.76  & RC02 \\
Si\,{\sc i}  & 6243.82 & 5.61 &  $-$1.27  & IK01 \\
Si\,{\sc i}  & 6244.48 & 5.61 &  $-$1.27  & IK01 \\
Si\,{\sc i}  & 6721.84 & 5.86 &  $-$0.94  & RC02 \\
  &    &    &    &  \\
Ca\,{\sc i}  & 5260.39 & 2.52 &  $-$1.72  & LUCK \\
Ca\,{\sc i}  & 5261.71 & 2.52 &  $-$0.58  & LUCK \\
Ca\,{\sc i}  & 5512.99 & 2.93 &  $-$0.45  & LUCK \\
Ca\,{\sc i}  & 5581.98 & 2.52 &  $-$0.56  & LUCK \\
Ca\,{\sc i}  & 5588.76 & 2.52 & 0.36 & LUCK \\
Ca\,{\sc i}  & 5590.13 & 2.52 &  $-$0.57  & LUCK \\
Ca\,{\sc i}  & 5601.29 & 2.52 &  $-$0.52  & LUCK \\
Ca\,{\sc i}  & 5857.46 & 2.93 & 0.23 & RC02 \\
Ca\,{\sc i}  & 6102.73 & 1.88 &  $-$0.79  & LUCK \\
Ca\,{\sc i}  & 6122.23 & 1.89 &  $-$0.32  & LUCK \\
Ca\,{\sc i}  & 6161.29 & 2.52 &  $-$1.27  & IK01 \\
Ca\,{\sc i}  & 6162.18 & 1.90 &  $-$0.09  & RC02 \\
Ca\,{\sc i}  & 6166.44 & 2.52 &  $-$1.14  & R03 \\
Ca\,{\sc i}  & 6169.04 & 2.52 &  $-$0.80  & R03 \\
Ca\,{\sc i}  & 6169.56 & 2.52 &  $-$0.48  & IK01 \\
Ca\,{\sc i}  & 6439.08 & 2.52 & 0.39 & LUCK \\
Ca\,{\sc i}  & 6455.60 & 2.52 &  $-$1.29  & IK01 \\
Ca\,{\sc i}  & 6471.67 & 2.52 &  $-$0.69  & IK01 \\
Ca\,{\sc i}  & 6493.79 & 2.52 &  $-$0.11  & LUCK \\
Ca\,{\sc i}  & 6499.65 & 2.52 &  $-$0.82  & IK01 \\
  &    &    &    &  \\
Sc\,{\sc ii}  & 5526.82 & 1.77 & 0.13 & PN00 \\
Sc\,{\sc ii}  & 5657.88 & 1.51 &  $-$0.50  & PN00 \\
Sc\,{\sc ii}  & 5667.15 & 1.50 &  $-$1.24  & PN00 \\
Sc\,{\sc ii}  & 5669.04 & 1.50 &  $-$1.12  & PN00 \\
Sc\,{\sc ii}  & 6245.61 & 1.51 &  $-$0.98  & PN00 \\
Sc\,{\sc ii}  & 6604.60 & 1.36 &  $-$1.48  & PN00 \\
  &    &    &    &  \\
Ti\,{\sc i}  & 5173.75 & 0.00 &  $-$1.12  & KB95 \\
Ti\,{\sc i}  & 5192.98 & 0.02 &  $-$1.01  & KB95 \\
Ti\,{\sc i}  & 5210.39 & 0.05 &  $-$0.88  & KB95 \\
Ti\,{\sc i}  & 5648.57 & 2.49 &  $-$0.25  & RC02 \\
Ti\,{\sc i}  & 5662.16 & 2.32 &  $-$0.11  & RC02 \\
Ti\,{\sc i}  & 5679.94 & 2.47 &  $-$0.58  & RC02 \\
Ti\,{\sc i}  & 5689.49 & 2.30 &  $-$0.47  & RC02 \\
Ti\,{\sc i}  & 5702.69 & 2.29 &  $-$0.57  & RC02 \\
Ti\,{\sc i}  & 5713.92 & 2.29 &  $-$0.84  & RC02 \\
Ti\,{\sc i}  & 5716.46 & 2.30 &  $-$0.70  & RC02 \\
Ti\,{\sc i}  & 5720.48 & 2.29 &  $-$0.90  & RC02 \\
Ti\,{\sc i}  & 5739.46 & 2.25 &  $-$0.60  & RC02 \\
Ti\,{\sc i}  & 5739.98 & 2.24 &  $-$0.67  & RC02 \\
Ti\,{\sc i}  & 5766.33 & 3.29 & 0.36 & RC02 \\
Ti\,{\sc i}  & 5866.46 & 1.07 &  $-$0.84  & RC02 \\
Ti\,{\sc i}  & 5880.27 & 1.05 &  $-$2.05  & RC02 \\
\hline
\end{tabular}
\end{table}

\addtocounter{table}{-1}

\begin{table} 
\caption{Atomic line list}
\begin{tabular}{lccrc}
\hline
Species &
Wavelength (\AA) &
EP (eV) &
log $gf$ &
Ref.$^{\mathrm{a}}$  \\
\hline
Ti\,{\sc i}  & 5903.32 & 1.07 &  $-$2.14  & RC02 \\
Ti\,{\sc i}  & 5922.11 & 1.05 &  $-$1.47  & IK01 \\
Ti\,{\sc i}  & 5937.81 & 1.07 &  $-$1.89  & RC02 \\
Ti\,{\sc i}  & 5941.75 & 1.05 &  $-$1.52  & RC02 \\
Ti\,{\sc i}  & 5953.16 & 1.89 &  $-$0.33  & RC02 \\
Ti\,{\sc i}  & 5965.83 & 1.88 &  $-$0.41  & IK01 \\
Ti\,{\sc i}  & 5978.54 & 1.87 &  $-$0.50  & IK01 \\
Ti\,{\sc i}  & 5999.68 & 2.17 &  $-$0.73  & RC02 \\
Ti\,{\sc i}  & 6091.17 & 2.27 &  $-$0.42  & RC02 \\
Ti\,{\sc i}  & 6092.80 & 1.89 &  $-$1.38  & RC02 \\
Ti\,{\sc i}  & 6126.22 & 1.07 &  $-$1.42  & IK01 \\
Ti\,{\sc i}  & 6146.22 & 1.87 &  $-$1.47  & RC02 \\
Ti\,{\sc i}  & 6186.15 & 2.17 &  $-$1.15  & RC02 \\
Ti\,{\sc i}  & 6258.10 & 1.44 &  $-$0.36  & RC02 \\
Ti\,{\sc i}  & 6258.71 & 1.46 &  $-$0.24  & RC02 \\
Ti\,{\sc i}  & 6261.10 & 1.43 &  $-$0.48  & RC02 \\
Ti\,{\sc i}  & 6303.76 & 1.44 &  $-$1.57  & IK01 \\
Ti\,{\sc i}  & 6312.24 & 1.46 &  $-$1.55  & IK01 \\
Ti\,{\sc i}  & 6497.69 & 1.44 &  $-$1.93  & RC02 \\
Ti\,{\sc i}  & 6508.14 & 1.43 &  $-$1.98  & RC02 \\
Ti\,{\sc i}  & 6554.22 & 1.44 &  $-$1.22  & RC02 \\
Ti\,{\sc i}  & 6716.68 & 2.49 &  $-$1.04  & RC02 \\
Ti\,{\sc i}  & 6743.12 & 0.90 &  $-$1.63  & RC02 \\
  &    &    &    &  \\
Ti\,{\sc ii}  & 5154.08 & 1.57 &  $-$1.92  & KB95 \\
Ti\,{\sc ii}  & 5185.91 & 1.89 &  $-$1.35  & KB95 \\
Ti\,{\sc ii}  & 5336.79 & 1.58 &  $-$1.69  & KB95 \\
Ti\,{\sc ii}  & 5381.03 & 1.59 &  $-$2.08  & KB95 \\
  &    &    &    &  \\
V\,{\sc i}  & 5727.06 & 1.08 &  $-$0.01  & R03 \\
V\,{\sc i}  & 6090.22 & 1.08 &  $-$0.06  & R03 \\
V\,{\sc i}  & 6216.36 & 0.28 &  $-$1.29  & PN00 \\
V\,{\sc i}  & 6251.82 & 0.29 &  $-$1.34  & PN00 \\
V\,{\sc i}  & 6274.64 & 0.27 &  $-$1.67  & PN00 \\
V\,{\sc i}  & 6504.16 & 1.18 &  $-$1.23  & PN00 \\
  &    &    &    &  \\
Mn\,{\sc i}  & 5537.74 & 2.19 &  $-$2.02  & PN00 \\
Mn\,{\sc i}  & 6013.53 & 3.07 &  $-$0.25  & PN00 \\
Mn\,{\sc i}  & 6016.67 & 3.08 &  $-$0.22  & PN00 \\
Mn\,{\sc i}  & 6021.80 & 3.07 & 0.03 & PN00 \\
  &    &    &    &  \\
Co\,{\sc i}  & 5342.71 & 4.02 & 0.54 & PN00 \\
Co\,{\sc i}  & 5352.05 & 3.58 & 0.06 & PN00 \\
Co\,{\sc i}  & 5530.79 & 1.71 &  $-$2.06  & PN00 \\
Co\,{\sc i}  & 6455.03 & 3.63 &  $-$0.25  & PN00 \\
Co\,{\sc i}  & 6632.45 & 2.28 &  $-$2.00  & PN00 \\
  &    &    &    &  \\
Ni\,{\sc i}  & 5578.73 & 1.68 &  $-$2.64  & KB95 \\
Ni\,{\sc i}  & 5682.20 & 4.10 &  $-$0.47  & RC02 \\
Ni\,{\sc i}  & 5748.35 & 1.68 &  $-$3.26  & RC02 \\
Ni\,{\sc i}  & 5892.88 & 1.99 &  $-$2.34  & RC02 \\
Ni\,{\sc i}  & 6007.31 & 1.68 &  $-$3.34  & RC02 \\
Ni\,{\sc i}  & 6086.28 & 4.26 &  $-$0.52  & RC02 \\
Ni\,{\sc i}  & 6108.12 & 1.68 &  $-$2.45  & KB95 \\
Ni\,{\sc i}  & 6175.37 & 4.09 &  $-$0.53  & RC02 \\
Ni\,{\sc i}  & 6176.82 & 4.09 &  $-$0.53  & RC02 \\
\hline
\end{tabular}
\end{table}

\addtocounter{table}{-1}

\begin{table} 
\caption{Atomic line list}
\begin{tabular}{lccrc}
\hline
Species &
Wavelength (\AA) &
EP (eV) &
log $gf$ &
Ref.$^{\mathrm{a}}$  \\
\hline
Ni\,{\sc i}  & 6177.25 & 1.83 &  $-$3.50  & RC02 \\
Ni\,{\sc i}  & 6186.71 & 4.10 &  $-$0.97  & RC02 \\
Ni\,{\sc i}  & 6204.60 & 4.09 &  $-$1.14  & RC02 \\
Ni\,{\sc i}  & 6223.99 & 4.10 &  $-$0.99  & IK01 \\
Ni\,{\sc i}  & 6322.17 & 4.15 &  $-$1.17  & RC02 \\
Ni\,{\sc i}  & 6360.82 & 4.17 &  $-$1.15  & RC02 \\
Ni\,{\sc i}  & 6370.35 & 3.54 &  $-$1.94  & RC02 \\
Ni\,{\sc i}  & 6378.26 & 4.15 &  $-$0.89  & RC02 \\
Ni\,{\sc i}  & 6598.60 & 4.23 &  $-$0.98  & RC02 \\
Ni\,{\sc i}  & 6635.12 & 4.42 &  $-$0.83  & RC02 \\
Ni\,{\sc i}  & 6643.64 & 1.68 &  $-$2.30  & RC02 \\
  &    &    &    &  \\
Cu\,{\sc i}  & 5105.54 & 1.39 &  $-$1.52  & SS03 \\
  &    &    &    &  \\
Y\,{\sc ii}  & 5123.22 & 0.99 &  $-$0.83  & PN00 \\
Y\,{\sc ii}  & 5200.42 & 0.99 &  $-$0.57  & PN00 \\
Y\,{\sc ii}  & 5509.91 & 0.99 &  $-$1.01  & KB95 \\
Y\,{\sc ii}  & 5544.61 & 1.74 &  $-$1.08  & RC02 \\
  &    &    &    &  \\
Zr\,{\sc i}  & 6127.44 & 0.15 &  $-$1.06  & RC02 \\
Zr\,{\sc i}  & 6134.55 & 0.00 &  $-$1.28  & RC02 \\
Zr\,{\sc i}  & 6143.20 & 0.07 &  $-$1.10  & RC02 \\
  &    &    &    &  \\
Zr\,{\sc ii}  & 5112.27 & 1.67 &  $-$0.59  & KB95 \\
  &    &    &    &  \\
Ba\,{\sc ii}  & 5853.64 & 0.60 &  $-$1.01  & PN00 \\
Ba\,{\sc ii}  & 6141.73 & 0.70 &  $-$0.08  & PN00 \\
Ba\,{\sc ii}  & 6496.91 & 0.60 &  $-$0.38  & PN00 \\
  &    &    &    &  \\
La\,{\sc ii}  & 5303.53 & 0.32 &  $-$1.35  & LB01 \\
La\,{\sc ii}  & 6390.49 & 0.30 &  $-$1.41  & LB01 \\
  &    &    &    &  \\
Ce\,{\sc ii}  & 4943.45 & 1.20 &  $-$0.11  & LUCK \\
Ce\,{\sc ii}  & 5274.24 & 1.04 &  $-$0.32  & LUCK \\
Ce\,{\sc ii}  & 5472.30 & 1.24 &  $-$0.18  & LUCK \\
Ce\,{\sc ii}  & 5512.06 & 1.00 & 0.29 & LUCK \\
  &    &    &    &  \\
Nd\,{\sc ii}  & 4959.12 & 0.06 &  $-$0.80  & DL03 \\
Nd\,{\sc ii}  & 5092.79 & 0.38 &  $-$0.61  & DL03 \\
Nd\,{\sc ii}  & 5212.36 & 0.20 &  $-$0.96  & DL03 \\
Nd\,{\sc ii}  & 5234.19 & 0.55 &  $-$0.51  & DL03 \\
Nd\,{\sc ii}  & 5249.58 & 0.98 & 0.20 & DL03 \\
Nd\,{\sc ii}  & 5255.51 & 0.20 &  $-$0.67  & DL03 \\
Nd\,{\sc ii}  & 5293.16 & 0.82 & 0.10 & DL03 \\
Nd\,{\sc ii}  & 5319.81 & 0.55 &  $-$0.14  & DL03 \\
  &    &    &    &  \\
Eu\,{\sc ii}  & 6645.13 & 1.37 & 0.20 & LW01 \\
\hline
\end{tabular}

\begin{list}{}{}
\item[$^{\mathrm{a}}$] References for the $gf$ values
\item[DL03] -- \citet{nd}
\item[IK01] -- \citet{M5}
\item[KB95] -- \citet{kurucz95}
\item[LW01] -- \citet{eu}
\item[LB01] -- \citet{la}
\item[LUCK] -- Luck (2003, private communication)
\item[PN00] -- \citet{prochaska00}
\item[RC02] -- \citet{ramirez02}
\item[SS03] -- \citet{simmerer03}
\end{list}

\end{table}

%% file: 2916.t3.tex
\begin{table*}
\centering
\caption{Elemental abundances for program stars (O-Mn) \label{tab:ab}}
\begin{tabular}{lrrrrrrrrrr}
\hline
Name1 &
[O/Fe] & 
[Na/Fe] & 
[Mg/Fe] & 
[Al/Fe] & 
[Si/Fe] & 
[Ca/Fe] & 
[Sc/Fe] & 
[Ti/Fe] & 
[V/Fe] & 
[Mn/Fe]
\\
\hline
NGC6752-mg0 & 0.17 & 0.67 & 0.48 & 1.08 & 0.38 & 0.17 & 0.01 & 0.13 & $-$0.32 & $-$0.43 \\
NGC6752-mg1 & 0.46 & 0.38 & 0.46 & 0.82 & 0.41 & 0.03 & $-$0.04 & 0.05 & $-$0.45 & $-$0.47 \\
NGC6752-mg2 & 0.55 & 0.19 & 0.47 & 0.77 & 0.43 & 0.09 & 0.01 & 0.09 & $-$0.43 & $-$0.47 \\
NGC6752-mg3 & 0.47 & 0.22 & 0.50 & 0.77 & 0.32 & 0.19 & $-$0.06 & 0.12 & $-$0.32 & $-$0.41 \\
NGC6752-mg4 & 0.38 & 0.29 & 0.47 & 0.90 & 0.34 & 0.18 & $-$0.01 & 0.12 & $-$0.33 & $-$0.41 \\
NGC6752-mg5 & 0.42 & 0.32 & 0.51 & 0.74 & 0.33 & 0.26 & $-$0.02 & 0.16 & $-$0.27 & $-$0.39 \\
NGC6752-mg6 & 0.60 & 0.13 & 0.54 & 0.57 & 0.34 & 0.28 & $-$0.18 & 0.18 & $-$0.24 & $-$0.38 \\
NGC6752-mg8 & 0.40 & 0.34 & 0.50 & 0.74 & 0.32 & 0.22 & $-$0.01 & 0.09 & $-$0.36 & $-$0.48 \\
NGC6752-mg9 & 0.47 & 0.28 & 0.51 & 0.77 & 0.31 & 0.29 & $-$0.05 & 0.18 & $-$0.18 & $-$0.39 \\
NGC6752-mg10 & 0.44 & 0.28 & 0.51 & 0.78 & 0.29 & 0.28 & 0.00 & 0.12 & $-$0.27 & $-$0.41 \\
NGC6752-mg12 & 0.66 & $-$0.09 & 0.48 & 0.09 & 0.27 & 0.30 & $-$0.04 & 0.12 & $-$0.26 & $-$0.42 \\
NGC6752-mg15 & 0.40 & 0.31 & 0.52 & 0.72 & 0.30 & 0.33 & $-$0.04 & 0.20 & $-$0.14 & $-$0.34 \\
NGC6752-mg18 & 0.46 & 0.19 & 0.49 & 0.59 & 0.31 & 0.36 & $-$0.07 & 0.20 & $-$0.19 & $-$0.36 \\
NGC6752-mg21 & 0.01 & 0.57 & 0.42 & 1.18 & 0.35 & 0.38 & 0.01 & 0.17 & $-$0.14 & $-$0.36 \\
NGC6752-mg22 & 0.19 & 0.63 & 0.49 & 0.99 & 0.35 & 0.37 & $-$0.05 & 0.17 & $-$0.15 & $-$0.42 \\
NGC6752-mg24 & 0.65 & $-$0.09 & 0.50 & 0.12 & 0.26 & 0.31 & $-$0.01 & 0.16 & $-$0.22 & $-$0.39 \\
NGC6752-mg25 & 0.59 & 0.14 & 0.53 & 0.51 & 0.33 & 0.38 & $-$0.04 & 0.19 & $-$0.14 & $-$0.39 \\
NGC6752-0 & $-$0.15 & 0.55 & 0.24 & 1.33 & 0.41 & 0.23 & $-$0.02 & 0.11 & $-$0.33 & $-$0.45 \\
NGC6752-1 & 0.57 & 0.08 & 0.52 & 0.32 & 0.35 & 0.23 & $-$0.14 & 0.13 & $-$0.27 & $-$0.44 \\
NGC6752-2 & $-$0.09 & 0.60 & 0.39 & 1.18 & 0.32 & 0.26 & $-$0.10 & 0.19 & $-$0.26 & $-$0.44 \\
NGC6752-3 & 0.70 & $-$0.04 & 0.52 & 0.22 & 0.31 & 0.21 & $-$0.03 & 0.17 & $-$0.29 & $-$0.55 \\
NGC6752-4 & $-$0.04 & 0.61 & 0.39 & 1.20 & 0.37 & 0.27 & $-$0.06 & 0.15 & $-$0.29 & $-$0.47 \\
NGC6752-6 & 0.09 & 0.54 & 0.46 & 0.96 & 0.27 & 0.25 & $-$0.06 & 0.12 & $-$0.19 & $-$0.50 \\
NGC6752-7 & 0.90 & 0.02 & 0.55 & 0.19 & 0.11 & 0.01 & $-$0.02 & $-$0.04 & $-$0.73 & $-$0.75 \\
NGC6752-8 & 0.66 & $-$0.01 & 0.54 & 0.48 & 0.30 & 0.23 & $-$0.07 & 0.14 & $-$0.47 & $-$0.56 \\
NGC6752-9 & 0.65 & $-$0.02 & 0.52 & 0.13 & 0.32 & 0.21 & $-$0.07 & 0.10 & $-$0.35 & $-$0.48 \\
NGC6752-10 & $-$0.02 & 0.65 & 0.43 & 1.06 & 0.36 & 0.24 & $-$0.04 & 0.16 & $-$0.35 & $-$0.45 \\
NGC6752-11 & 0.37 & 0.35 & 0.47 & 0.90 & 0.24 & 0.19 & $-$0.02 & 0.08 & $-$0.35 & $-$0.54 \\
NGC6752-12 & 0.29 & 0.27 & 0.50 & 0.41 & 0.35 & 0.23 & 0.01 & 0.14 & $-$0.35 & $-$0.49 \\
NGC6752-15 & 0.65 & $-$0.10 & 0.51 & 0.58 & 0.36 & 0.22 & $-$0.05 & 0.11 & $-$0.19 & $-$0.39 \\
NGC6752-16 & 0.09 & 0.36 & 0.48 & 0.83 & 0.40 & 0.22 & $-$0.04 & 0.17 & $-$0.29 & $-$0.48 \\
NGC6752-19 & 0.29 & 0.22 & 0.48 & 0.59 & 0.29 & 0.21 & $-$0.04 & 0.13 & $-$0.60 & $-$0.52 \\
NGC6752-20 & 0.08 & 0.67 & 0.37 & 1.15 & 0.35 & 0.26 & $-$0.05 & 0.16 & $-$0.16 & $-$0.44 \\
NGC6752-21 & 0.49 & 0.29 & 0.46 & 0.63 & 0.31 & 0.23 & 0.00 & 0.15 & $-$0.22 & $-$0.51 \\
NGC6752-23 & 0.11 & 0.59 & 0.34 & 1.25 & 0.36 & 0.24 & 0.00 & 0.12 & $-$0.29 & $-$0.51 \\
NGC6752-24 & 0.56 & 0.01 & 0.49 & 0.36 & 0.20 & 0.18 & $-$0.02 & 0.08 & $-$0.33 & $-$0.53 \\
NGC6752-29 & 0.51 & $-$0.07 & 0.48 & 0.35 & 0.25 & 0.21 & $-$0.07 & 0.05 & $-$0.24 & $-$0.50 \\
NGC6752-30 & 0.61 & 0.15 & 0.49 & 0.56 & 0.32 & 0.28 & $-$0.04 & 0.16 & $-$0.17 & $-$0.45 \\
\hline
\end{tabular}

\end{table*}

%% file: 2916.t4.tex
\begin{table*}
\centering
\caption{Elemental abundances for program stars (Co-Eu) \label{tab:ab2}}
\begin{tabular}{lrrrrrrrrrr}
\hline
Name1 &
[Co/Fe] & 
[Ni/Fe] & 
[Cu/Fe] & 
[Y/Fe] & 
[Zr/Fe] & 
[Ba/Fe] & 
[La/Fe] & 
[Ce/Fe] & 
[Nd/Fe] & 
[Eu/Fe]
\\
\hline
NGC6752-mg0 & $-$0.02 & $-$0.10 & $-$0.61 & 0.07 & 0.23 & \ldots & 0.07 & 0.20 & 0.24 & 0.32 \\
NGC6752-mg1 & $-$0.02 & $-$0.13 & $-$0.71 & 0.04 & 0.19 & \ldots & 0.05 & 0.23 & 0.22 & 0.34 \\
NGC6752-mg2 & 0.00 & $-$0.12 & $-$0.66 & 0.10 & 0.20 & \ldots & 0.08 & 0.25 & 0.26 & 0.35 \\
NGC6752-mg3 & $-$0.01 & $-$0.10 & $-$0.59 & 0.06 & 0.20 & $-$0.33 & 0.05 & 0.23 & 0.17 & 0.34 \\
NGC6752-mg4 & 0.00 & $-$0.11 & $-$0.61 & 0.06 & 0.19 & $-$0.35 & 0.07 & 0.24 & 0.29 & 0.35 \\
NGC6752-mg5 & 0.02 & $-$0.08 & $-$0.60 & 0.01 & 0.19 & $-$0.15 & 0.06 & 0.28 & 0.27 & 0.33 \\
NGC6752-mg6 & 0.04 & $-$0.07 & $-$0.53 & 0.07 & 0.27 & $-$0.23 & 0.09 & 0.30 & 0.13 & 0.36 \\
NGC6752-mg8 & $-$0.03 & $-$0.11 & $-$0.60 & $-$0.12 & 0.17 & $-$0.31 & 0.17 & 0.17 & 0.24 & 0.30 \\
NGC6752-mg9 & 0.01 & $-$0.07 & $-$0.55 & $-$0.05 & 0.22 & $-$0.13 & 0.10 & 0.25 & 0.24 & 0.30 \\
NGC6752-mg10 & $-$0.02 & $-$0.09 & $-$0.57 & 0.02 & 0.20 & $-$0.02 & 0.07 & 0.21 & 0.23 & 0.36 \\
NGC6752-mg12 & 0.01 & $-$0.09 & $-$0.55 & $-$0.05 & 0.17 & $-$0.03 & 0.10 & 0.23 & 0.24 & 0.34 \\
NGC6752-mg15 & 0.06 & $-$0.04 & $-$0.51 & $-$0.05 & 0.30 & $-$0.11 & 0.12 & 0.25 & 0.19 & 0.23 \\
NGC6752-mg18 & 0.06 & $-$0.03 & $-$0.53 & $-$0.04 & 0.26 & $-$0.08 & 0.12 & 0.29 & 0.12 & 0.36 \\
NGC6752-mg21 & $-$0.13 & $-$0.05 & $-$0.53 & 0.01 & 0.27 & $-$0.05 & 0.17 & 0.28 & 0.28 & 0.37 \\
NGC6752-mg22 & 0.03 & $-$0.02 & $-$0.53 & 0.00 & 0.31 & 0.01 & 0.16 & 0.25 & 0.21 & 0.31 \\
NGC6752-mg24 & 0.04 & $-$0.05 & $-$0.53 & $-$0.19 & 0.33 & $-$0.25 & 0.13 & 0.24 & 0.18 & 0.35 \\
NGC6752-mg25 & 0.06 & $-$0.03 & $-$0.55 & $-$0.03 & 0.37 & 0.03 & 0.16 & 0.30 & 0.20 & 0.36 \\
NGC6752-0 & $-$0.01 & $-$0.04 & $-$0.62 & 0.02 & 0.28 & 0.17 & \ldots & 0.30 & 0.20 & 0.31 \\
NGC6752-1 & $-$0.04 & $-$0.01 & $-$0.60 & 0.02 & 0.08 & 0.12 & \ldots & 0.33 & 0.18 & 0.33 \\
NGC6752-2 & $-$0.05 & 0.00 & $-$0.61 & 0.01 & 0.19 & 0.09 & \ldots & 0.36 & 0.12 & 0.45 \\
NGC6752-3 & $-$0.18 & $-$0.01 & $-$0.66 & $-$0.11 & $-$0.05 & $-$0.14 & \ldots & 0.32 & 0.17 & 0.33 \\
NGC6752-4 & $-$0.18 & $-$0.04 & $-$0.63 & 0.00 & 0.00 & 0.10 & \ldots & 0.25 & 0.25 & 0.50 \\
NGC6752-6 & $-$0.10 & $-$0.07 & $-$0.64 & $-$0.02 & 0.23 & 0.00 & \ldots & 0.34 & 0.24 & 0.35 \\
NGC6752-7 & $-$0.14 & $-$0.23 & $-$0.86 & $-$0.14 & $-$0.18 & $-$0.24 & \ldots & 0.15 & 0.30 & 0.25 \\
NGC6752-8 & $-$0.02 & 0.02 & $-$0.68 & $-$0.04 & 0.20 & 0.10 & \ldots & 0.39 & 0.25 & 0.46 \\
NGC6752-9 & 0.02 & $-$0.05 & $-$0.63 & $-$0.05 & 0.09 & 0.00 & \ldots & 0.42 & 0.26 & 0.33 \\
NGC6752-10 & $-$0.14 & $-$0.02 & $-$0.61 & 0.02 & 0.17 & 0.08 & \ldots & 0.27 & 0.26 & 0.35 \\
NGC6752-11 & $-$0.14 & $-$0.06 & $-$0.66 & $-$0.04 & 0.09 & 0.04 & \ldots & 0.30 & 0.28 & 0.16 \\
NGC6752-12 & $-$0.14 & $-$0.01 & $-$0.64 & $-$0.11 & 0.08 & 0.00 & \ldots & 0.19 & 0.25 & 0.31 \\
NGC6752-15 & $-$0.14 & $-$0.05 & $-$0.65 & $-$0.14 & $-$0.10 & $-$0.13 & \ldots & 0.34 & 0.22 & 0.13 \\
NGC6752-16 & $-$0.02 & 0.04 & $-$0.56 & $-$0.03 & 0.22 & $-$0.05 & \ldots & 0.23 & 0.23 & 0.10 \\
NGC6752-19 & $-$0.02 & $-$0.03 & $-$0.66 & $-$0.10 & 0.18 & $-$0.05 & \ldots & 0.26 & 0.22 & 0.21 \\
NGC6752-20 & $-$0.02 & 0.01 & $-$0.59 & 0.02 & 0.21 & 0.05 & \ldots & 0.36 & 0.25 & 0.23 \\
NGC6752-21 & $-$0.02 & $-$0.03 & $-$0.61 & $-$0.03 & 0.18 & $-$0.08 & \ldots & 0.25 & 0.25 & 0.33 \\
NGC6752-23 & $-$0.02 & $-$0.01 & $-$0.65 & $-$0.01 & 0.17 & 0.01 & \ldots & 0.19 & 0.25 & 0.17 \\
NGC6752-24 & 0.03 & $-$0.06 & $-$0.70 & $-$0.15 & 0.04 & $-$0.15 & \ldots & 0.18 & 0.25 & 0.19 \\
NGC6752-29 & 0.11 & $-$0.08 & $-$0.71 & $-$0.09 & 0.11 & $-$0.13 & \ldots & 0.24 & 0.21 & 0.40 \\
NGC6752-30 & 0.11 & $-$0.07 & $-$0.57 & 0.01 & 0.16 & $-$0.06 & \ldots & 0.18 & 0.24 & 0.35 \\
\hline
\end{tabular}

\end{table*}

%% file: 2916.t5.tex
\begin{table*}
\caption{Abundance dependences on model parameters
\label{tab:error}}
\begin{tabular}{lrrrrrr}
\hline
 &
\multicolumn{3}{c}{NGC6752-mg6$^{\mathrm{a}}$} &
\multicolumn{3}{c}{NGC6752-15$^{\mathrm{b}}$} 
\\
Abundance &
\teff~+ 30&
log g + 0.1 &
$\xi_t$ + 0.1 &
\teff~+ 30&
log g + 0.1 &
$\xi_t$ + 0.1
\\
\hline
{\rm [Fe/H]} & 0.01 & 0.01 & $-$0.02 & 0.03 & $-$0.01 & $-$0.03 \\
{\rm [O/Fe]} & 0.01 & 0.04 & $-$0.01 & $-$0.02 & 0.05 & 0.03 \\
{\rm [Na/Fe]} & 0.03 & $-$0.02 & $-$0.01 & $-$0.01 & $-$0.01 & 0.03 \\
{\rm [Mg/Fe]} & 0.02 & $-$0.02 & $-$0.03 & $-$0.01 & $-$0.02 & 0.01 \\
{\rm [Al/Fe]} & 0.03 & $-$0.01 & $-$0.01 & $-$0.01 & $-$0.01 & 0.02 \\
{\rm [Si/Fe]} & $-$0.01 & 0.02 & $-$0.01 & $-$0.02 & 0.01 & 0.02 \\
{\rm [Ca/Fe]} & 0.04 & $-$0.01 & $-$0.04 & $-$0.01 & $-$0.01 & $-$0.01 \\
{\rm [Sc/Fe]} & $-$0.01 & 0.04 & $-$0.02 & $-$0.03 & 0.04 & 0.01 \\
{\rm [Ti/Fe]} & 0.05 & $-$0.01 & $-$0.02 & 0.01 & 0.01 & $-$0.02 \\
{\rm [V/Fe]} & 0.07 & $-$0.01 & $-$0.01 & 0.02 & $-$0.01 & 0.03 \\
{\rm [Mn/Fe]} & 0.04 & $-$0.01 & $-$0.01 & $-$0.01 & $-$0.01 & 0.02 \\
{\rm [Co/Fe]} & 0.02 & $-$0.01 & $-$0.01 & 0.02 & 0.01 & 0.04 \\
{\rm [Ni/Fe]} & 0.02 & 0.02 & $-$0.02 & $-$0.01 & 0.01 & 0.02 \\
{\rm [Cu/Fe]} & 0.02 & $-$0.01 & $-$0.02 & \ldots & \ldots & \ldots \\
{\rm [Y/Fe]} & 0.01 & 0.03 & $-$0.02 & $-$0.03 & 0.04 & 0.02 \\
{\rm [Zr/Fe]} & 0.08 & 0.01 & $-$0.01 & \ldots & \ldots & \ldots \\
{\rm [Ba/Fe]} & 0.01 & 0.04 & $-$0.07 & $-$0.01 & 0.04 & $-$0.04 \\
{\rm [La/Fe]} & $-$0.01 & 0.03 & 0.03 & \ldots & \ldots & \ldots \\
{\rm [Ce/Fe]} & 0.01 & 0.04 & $-$0.01 & $-$0.02 & 0.04 & 0.03 \\
{\rm [Nd/Fe]} & 0.01 & 0.03 & $-$0.02 & $-$0.02 & 0.04 & 0.02 \\
{\rm [Eu/Fe]} & $-$0.02 & 0.03 & $-$0.02 & $-$0.02 & 0.04 & 0.03 \\
\hline
\end{tabular}

\begin{list}{}{}
\item[$^{\mathrm{a}}$]NGC6752-mg6: \teff=4154 K, log g=0.68 cm s$^{-2}$, $\xi_t$=2.10 km s$^{-1}$
\item[$^{\mathrm{b}}$]NGC6752-15: \teff=4850 K, log g=2.19 cm s$^{-2}$, $\xi_t$=1.35 km s$^{-1}$
\end{list}

\end{table*}

%% file: 2916.t6.tex
\begin{table*}
\caption{Mean Abundances and comparison of predicted
and observed spread
\label{tab:pred}}
\begin{tabular}{lrcc}
\hline
{\rm [X/Fe]} &
{\rm Mean} &
$\sigma_{\rm predicted}$ &
$\sigma_{\rm observed}$
\\
\hline
{\rm [Fe/H]} & $-$1.61 & 0.03 & 0.02 \\
{\rm [O/Fe]} & 0.37 & 0.05 & 0.25 \\
{\rm [Na/Fe]} & 0.28 & 0.03 & 0.24 \\
{\rm [Mg/Fe]} & 0.47 & 0.03 & 0.06 \\
{\rm [Al/Fe]} & 0.72 & 0.03 & 0.34 \\
{\rm [Si/Fe]} & 0.33 & 0.03 & 0.05 \\
{\rm [Ca/Fe]} & 0.24 & 0.04 & 0.07 \\
{\rm [Sc/Fe]} & $-$0.04 & 0.04 & 0.04 \\
{\rm [Ti/Fe]} & 0.14 & 0.04 & 0.04 \\
{\rm [V/Fe]} & $-$0.28 & 0.05 & 0.10 \\
{\rm [Mn/Fe]} & $-$0.45 & 0.03 & 0.06 \\
{\rm [Co/Fe]} & $-$0.02 & 0.03 & 0.07 \\
{\rm [Ni/Fe]} & $-$0.05 & 0.03 & 0.04 \\
{\rm [Cu/Fe]} & $-$0.61 & 0.03 & 0.05 \\
{\rm [Y/Fe]} & $-$0.02 & 0.04 & 0.07 \\
{\rm [Zr/Fe]} & 0.18 & 0.08 & 0.10 \\
{\rm [Ba/Fe]} & $-$0.06 & 0.07 & 0.13 \\
{\rm [La/Fe]} & 0.10 & 0.04 & 0.04 \\
{\rm [Ce/Fe]} & 0.27 & 0.05 & 0.06 \\
{\rm [Nd/Fe]} & 0.22 & 0.04 & 0.04 \\
{\rm [Eu/Fe]} & 0.32 & 0.05 & 0.09 \\
\hline
\end{tabular}

Note. --- Star NGC6752-7 has been omitted due to its deviating [Fe/H].

\end{table*}

%% file: 2916.t7.tex
\begin{table*}
\caption{Abundance comparison with literature 
\label{tab:comp}}
\begin{tabular}{lrrrrr}
\hline
&
This Study &
ND95 &
\multicolumn{2}{c}{James04} &
CPS04
\\
&
&
&
Subgiants &
Dwarfs
\\
Species &
Mean ($\sigma$) &
Mean ($\sigma$) &
Mean ($\sigma$) &
Mean ($\sigma$) &
Mean ($\sigma$)
\\
\hline
{\rm [Si/Fe]} & 0.33 (0.05) & 0.26 (0.06) & \ldots & \ldots & \ldots \\
{\rm [Ca/Fe]} & 0.24 (0.07) & 0.40 (0.02) & \ldots & \ldots & 0.26 (0.08) \\
{\rm [Sc/Fe]} & $-$0.04 (0.04) & 0.00 (0.05) & \ldots & \ldots & \ldots \\
{\rm [Ti/Fe]} & 0.14 (0.04) & 0.15 (0.09) & \ldots & \ldots & 0.24 (0.18) \\
{\rm [V/Fe]} & $-$0.28 (0.10) & $-$0.01 (0.08) & \ldots & \ldots & \ldots \\
{\rm [Fe/H]} & $-$1.61 (0.02) & $-$1.52 (0.04) & $-$1.49 (0.07) & $-$1.48 (0.07) & $-$1.58 (0.16) \\
{\rm [Ni/Fe]} & $-$0.05 (0.04) & $-$0.16 (0.03) & \ldots & \ldots & $-$0.12 (0.12) \\
{\rm [Y/Fe]} & $-$0.02 (0.07) & $-$0.27 (0.09) & $-$0.01 (0.13) & $-$0.03 (0.11) & \ldots \\
{\rm [Zr/Fe]} & 0.18 (0.10) & 0.17 (0.06) & \ldots & \ldots & \ldots \\
{\rm [Ba/Fe]} & $-$0.06 (0.13) & 0.00 (0.13) & 0.25 (0.08) & 0.11 (0.09) & \ldots \\
{\rm [La/Fe]} & 0.10 (0.04) & $-$0.07 (0.07) & \ldots & \ldots & 0.13 (0.14) \\
{\rm [Nd/Fe]} & 0.22 (0.04) & $-$0.07 (0.10) & \ldots & \ldots & \ldots \\
{\rm [Eu/Fe]} & 0.32 (0.09) & $-$0.25 (0.06) & 0.40 (0.09) & 0.47 (0.08) & 0.55 (0.12) \\
\hline
\end{tabular}

Note. -- ND95 = \citealt{norris95}, CSP04 = \citealt{csp04}, and James04 = \citealt{james04}. Star NGC6752-7 has been omitted due to its deviating [Fe/H]. 

\end{table*}

%% file: 2916.t8.tex
\begin{table*}
\caption{Cluster metallicities and Abundance references
\label{tab:ref}}
\begin{tabular}{lrl}
\hline
NGC (Other) &
[Fe/H]$^{\mathrm{a}}$ &
Reference
\\
\hline
104 (47 Tuc) & $-$0.70 & \citet{brown92} \\
288 & $-$1.41 & \citet{shetrone00} \\
362 & $-$1.34 & \citet{shetrone00} \\
3201 & $-$1.56 & \citet{gonzalez98} \\
5272 (M3) & $-$1.50 & \citet{sneden04a} \\
5904 (M5) & $-$1.26 & \citet{M5,ramirez03} \\
6121 (M4) & $-$1.15 & \citet{M4} \\
6205 (M13) & $-$1.60 & \citet{kraft97,sneden04a} \\
6254 (M10) & $-$1.51 & \citet{kraft95} \\
6341 (M92) & $-$2.38 & \citet{shetrone96a,sneden00} \\
6397 & $-$2.02 & \citet{castilho00} \\
6752 & $-$1.61 & This study \\
6838 (M71) & $-$0.81 & \citet{ramirez02} \\
7078 (M15) & $-$2.42 & \citet{sneden97,sneden00} \\
---- (Pal 12) & $-$0.95 & \citet{cohen04} \\
\hline
\end{tabular}

\begin{list}{}{}
\item[$^{\mathrm{a}}$][Fe/H] values are from \citet{kraft03,kraft04}.
\end{list}


\end{table*}